\documentclass[aps,prd,nofootinbib,amssymb,eqsecnum,notitlepage]{revtex4-1}

\usepackage{bm}
\usepackage{amsmath,amsthm,amssymb}
\usepackage[dvipdfmx]{graphicx}
\usepackage{color}
\usepackage{amsfonts}
\usepackage[english]{babel}
\usepackage{hyperref}

\definecolor{MyBlue}{rgb}{0.15,0.15,0.70}

\hypersetup{colorlinks=true,citecolor=MyBlue,linkcolor=MyBlue,urlcolor=MyBlue}

\def\beq{\begin{equation}}
\def\eeq{\end{equation}}
\def\be{\begin{equation}}
\def\ee{\end{equation}}
\newcommand{\bea}{\begin{eqnarray}}
\newcommand{\eea}{\end{eqnarray}}
\def\bi{\begin{itemize}}
\def\ei{\end{itemize}}
\def\ba{\begin{array}}
\def\ea{\end{array}}
\def\bfig{\begin{figure}}
\def\efig{\end{figure}}
\def\d{\delta}

\def\B{{\cal B}}
\def\a{\alpha}
\def\b{\beta}
\def\f{F}

\def\aa{A_1}
\def\ab{A_2}
\def\ac{A_3}
\def\ad{A_4}
\def\ae{A_5}

\def\ba{\beta_1}

\def\A{{\cal A}}
\def\V{{\cal V}}
\def\R{{}^{(3)}\!R}
\def\D{D}

\newcommand{\quadac}{{\rm  quad}}
\newcommand{\aM}{\alpha_{\rm M}}
\newcommand{\aB}{\alpha_{\rm B}}
\newcommand{\aK}{\alpha_{\rm K}}
\newcommand{\aL}{\alpha_{\rm L}}
\newcommand{\aH}{\alpha_{\rm H}}
\newcommand{\aT}{\alpha_{\rm T}}
\newcommand{\bun}{\beta_1}
\newcommand{\bdeux}{\beta_2}
\newcommand{\btrois}{\beta_3}
\newcommand{\Az}{A}
\newcommand{\Bz}{B}

\newcommand{\CI}{{\cal C}_{\rm I}}
\newcommand{\CII}{{\cal C}_{\rm II}}

\newcommand{\Gfour}{G_4{}}
\newcommand{\Ffour}{F_4{}}

\newcommand{\2}{{(2)}}
\newcommand{\3}{{(3)}}

\newcommand\fX{{F_X}}
\newcommand\p{P}
\newcommand\q{Q}
\newcommand\h{A_3}
\newcommand\Am{{\cal A}}
\newcommand\M{{\cal M}}
\newcommand\GN{G_{\rm N}}

\newcommand\Q{Q}
\def\An{Q_*}
\def\tA{\hat Q} 

\begin{document}

\title{Dark Energy and Modified Gravity in Degenerate Higher-Order Scalar-Tensor (DHOST) theories: a review}
\author{David  Langlois}
%\email{langlois@apc.univ-paris7.fr}
\affiliation{Laboratoire Astroparticules et Cosmologie (APC), CNRS, Universit\'e Paris Diderot,\\
10 Rue Alice Domon et L\'eonie Duquet 75013 Paris, France}

\begin{abstract}
This article  reviews scalar-tensor theories characterized by a  Lagrangian that, despite the presence of second order derivatives,  contain a single scalar degree of freedom. These theories, known as Degenerate Higher-Order Scalar-Tensor (DHOST) theories, include  Horndeski and Beyond Horndeski theories. They  propagate a single scalar mode as a consequence of the degeneracy of their Lagrangian and, therefore, are  not plagued by an Ostrodradsky instability. They  have  been fully classified up to cubic order in second-order derivatives. The study of their  phenomenological consequences restricts  the subclass of DHOST theories that are compatible with observations. In cosmology, these theories can be described in the language of  the unified effective approach to dark energy and modified gravity. Compact objects in the context of DHOST theories are also discussed. 
\end{abstract}

\maketitle

\tableofcontents

\section{Introduction}

General Relativity (GR), which is now more than centennial,  has so far been  confirmed by observations, at least provided  a cosmological constant is added to the original theory in order to account for the present cosmological acceleration. The first direct observations of gravitational waves  are also in agreement with the main predictions of GR, paving the way for  refined tests of gravity in the strong field regime in the near future. 

Despite these observational successes, there have been numerous attempts to modify or extend general relativity, spurred by various motivations. One motivation is to replace invisible ingredients that appear to be  required by observations, such as dark energy or dark matter, by a modification of gravity on galactic or cosmological scales. Another motivation is to take into account, at least from an effective point of view, the corrections to general relativity that could resolve the singularities associated with black holes or with the early universe.
A more down-to-earth  motivation  is to test  general relativity quantitatively by  constructing a parametrized space of theories around GR,  which could be constrained by observations.

Because of their simplicity,  scalar-tensor theories have  played a prominent role in these attempts to go beyond GR. Many of the traditional scalar-tensor extensions of general relativity are described by a total action of  the  form 
 \beq
S=\frac{1}{16\pi \bar G}\int d^4x\sqrt{-g}\left[ F(\phi) R- Z(\phi) g^{\mu\nu}\partial_\mu\phi\partial_\nu\phi- 2U(\phi)\right]
+S_m\left[\psi_m, g_{\mu\nu}\right]\,,
\label{gJBD}
\eeq
which can be seen as generalizations  of the Brans-Dicke theories. 
The metric $g_{\mu\nu}$ in the above action corresponds to the physical (or Jordan) frame metric to which matter is minimally coupled, as indicated in the matter action $S_m$ which does not depend on $\phi$. Note that the general form (\ref{gJBD}) includes some of the most studied models in the cosmological context, such as
quintessence and  $f(R)$ theories  which can be recast as scalar-tensor theories. K-essence is an extension of the traditional form (\ref{gJBD}), where the Lagrangian is allowed to be a non-linear function of the kinetic term $X\equiv \nabla^\mu\phi\, \nabla_\mu \phi$.

More recently,  special attention has been devoted to scalar-tensor theories whose Lagrangians contain second-order derivatives of a scalar field, i.e. of the form
\beq
 {\cal L}(\nabla_\mu\!\nabla_\nu\phi, \nabla_\mu\phi, \phi\, ; g_{\mu\nu})\,.
 \eeq
Lagrangians of this type, which contain second-order time derivative $\ddot\phi$, are generically plagued by an instability due to the presence, in addition to the usual scalar mode and tensor modes, of an extra scalar degree of freedom (unless the higher order terms are restricted to be perturbative terms in the sense of low energy effective theories). This instability, known as Ostrogradsky instability, is related to a linear dependence of the Hamiltonian on one of the conjugate momenta~\cite{Ostrogradsky:1850fid,Woodard:2015zca}.

Until a few years ago, it was believed that only theories that yield second-order Euler-Lagrange equations, for both the scalar field and the metric, were free of this dangerous extra degree of freedom. Recently, it has been realized that this criterion is too restrictive and that there actually exists a much larger class of theories that satisfy this property, including theories  that lead to higher-order Euler-Lagrange equations. What characterizes these theories is not the order of their equation of motion but the degeneracy of their Lagrangian, which guarantees  that  only one scalar degree of freedom is present.  For this reason, these theories have been dubbed Degenerate Higher-Order Scalar-Tensor (DHOST) theories. The goal of this review is to present these theories and some of their phenomenological consequences.

The outline of this article is the following. In the next section, we briefly present the main scalar-tensor theories with higher-order derivatives that have been investigated in the last few years, from Horndeski to DHOST theories. Section \ref{section_degenerate} is devoted to the notion of degeneracy which is crucial to understand how the presence of an extra scalar degree of freedom can be avoided. The idea of degeneracy is then used in section \ref{section_DHOST} where quadratic DHOST theories are explicitly constructed. The consequences of DHOST theories in cosmology and for astrophysical objects are then discussed in sections \ref{section_cosmology} and \ref{section_gravity}, respectively. Section \ref{section_GW} concentrates on the status of DHOST theories after the observation of a binary neutron star merger via both gravitational waves and gamma-rays. Some other aspects and perspectives are discussed in the conclusion.

\section{Scalar-tensor theories with second order derivatives}
\begin{figure}[h]
\begin{center}
\includegraphics[height=5.5in,width=4in, angle=-90]{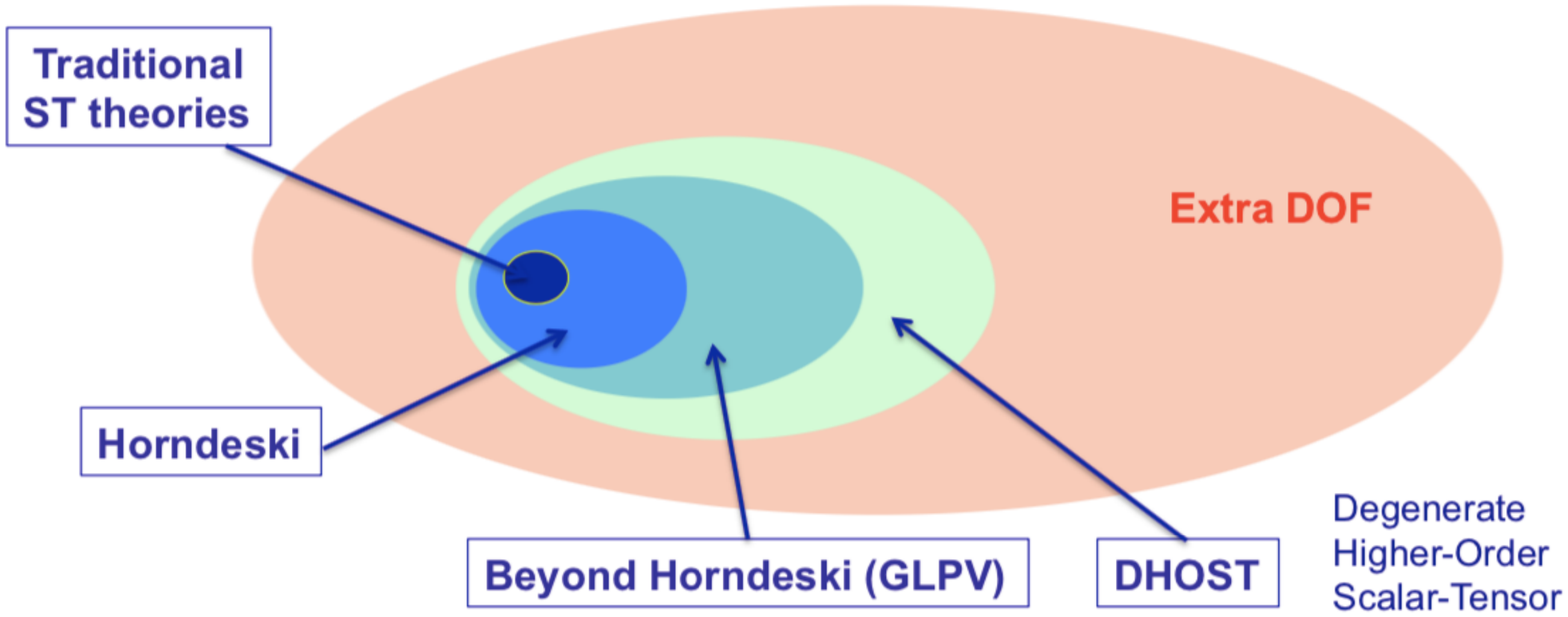}
\end{center}
\caption{\label{fig_ST}
\small Landscape of scalar-tensor theories.}
\end{figure}

In addition to the traditional scalar-tensor theories, whose Lagrangian depends at most on first order derivatives of the scalar field, the space of theories up to second order derivatives in their Lagrangian that contain a single scalar degree of freedom has been reappraised in the last few years. As a result,  the ``island''  of  safe theories has undergone an expansion  from Horndeski theories up to DHOST theories (see Fig. \ref{fig_ST}).

%%%%%%%%%%%%%%%%%%%%%%%%%%%%%%
%%%%%%%%%%%%%%%%%%%%%%%%%%%%%%
\subsection{Horndeski theories}
Scalar-tensor theories with a Lagrangian that contains second-order derivatives of a scalar field were constructed by Horndeski in 1974, with the specific  requirement that the Euler-Lagrange equations are at most second order~\cite{Horndeski:1974wa}.
In the modern presentation, the general action of Horndeski theories are usually written in terms of five elementary Lagrangians:
\begin{align}
L_2^{\rm H} & \equiv  G_2(\phi,X)\;, \qquad \qquad  
L_3^{\rm H}  \equiv  G_3(\phi, X) \, \Box \phi \;, \label{L3} \\
L_4^{\rm H} & \equiv G_4(\phi,X) \, {}^{(4)}\!R - 2 G_4{}_{,X}(\phi,X) (\Box \phi^2 - \phi^{ \mu \nu} \phi_{ \mu \nu})\;, \label{L4} \\
L_5^{\rm H} & \equiv G_5(\phi,X) \, {}^{(4)}\!G_{\mu \nu} \phi^{\mu \nu}  +  \frac13  G_5{}_{,X} (\phi,X) (\Box \phi^3 - 3 \, \Box \phi \, \phi_{\mu \nu}\phi^{\mu \nu} + 2 \, \phi_{\mu \nu}  \phi^{\mu \sigma} \phi^{\nu}_{\  \sigma})  \,, \label{L5}
\end{align}
with the notation
\beq
\phi_\mu\equiv \nabla_\mu\phi\,, \qquad \phi_{\mu\nu}\equiv \nabla_\nu\nabla_\mu\phi\,, \qquad X\equiv \phi^\mu\phi_\mu\,.
\eeq
Each of these Lagrangians depends on an arbitrary function of $\phi$ and $X$, denoted $G_A(\phi, X)$. Note that if one considers only $G_2(\phi,X)$ and $G_4(\phi)$, the Lagrangian does not contain any second order derivative and one recovers the general case of K-essence, mentioned in the introduction. 
Horndeski's work was mostly forgotten for many years, until it was resurrected in \cite{Charmousis:2011bf}. At that time, Horndeski theories had just been rediscovered  as a generalization  of the galileon models~\cite{Nicolis:2008in,Deffayet:2011gz,Kobayashi:2011nu}.

\subsection{Beyond Horndeski (GLPV) theories}
For a long time, it was believed that second-order Euler-Lagrange equations, as required by Horndeski,  were necessary  to have a single scalar degree of freedom. A first breach in this argument was provided in \cite{Zumalacarregui:2013pma}, where it was noticed  that (invertible)  field redefinitions of the metric  of the form~\cite{Bekenstein:1992pj}
\beq
\tilde g_{\mu\nu}=C(X, \phi) g_{\mu\nu}+D(X, \phi) \, \phi_\mu\, \phi_\nu\,
\eeq
in the Einstein-Hilbert action yields  a new Lagrangian leading to third-order equations of motion, the number of degrees of freedom being preserved in an invertible field redefinition (see e.g. \cite{Domenech:2015tca}).

Soon after,  following a different line of reasoning,  it was proposed in \cite{Gleyzes:2014dya,Gleyzes:2014qga} to extend the four elementary Horndeski Lagrangians (\ref{L3}-\ref{L5}) with the following two Lagrangians
\begin{align}
L_4^{\rm bH}& \equiv F_4(\phi,X) \epsilon^{\mu\nu\rho}_{\ \ \ \ \sigma}\, \epsilon^{\mu'\nu'\rho'\sigma}\phi_{\mu}\phi_{\mu'}\phi_{\nu\nu'}\phi_{\rho\rho'}\;, \label{L4bH} \\
L_5^{\rm bH}&\equiv F_5 (\phi,X) \epsilon^{\mu\nu\rho\sigma}\epsilon^{\mu'\nu'\rho'\sigma'}\phi_{\mu}\phi_{\mu'}\phi_{\nu \nu'}\phi_{\rho\rho'}\phi_{\sigma\sigma'} \label{L5bH}\,,
\end{align}
which depend on the two extra functions $F_4(\phi,X)$ and $F_5(\phi,X)$. 

Despite  leading to third order equations of motion, there exist theories  that propagate  a single scalar degree of freedom for appropriate combinations of the six  Lagrangians (\ref{L3}-\ref{L5}) and (\ref{L4bH}-\ref{L5bH}). It should be stressed that not all combinations of these six Lagrangians are allowed, but only those for which the total Lagrangian is degenerate (in the sense that will be defined precisely below), which means a special tuning of the functions $F_4$ or $F_5$ (see \cite{Langlois:2015cwa,Crisostomi:2016tcp,BenAchour:2016fzp} for details). 
This extended class of theories, which include Horndeski theories as a particular case, is usually denoted Beyond Horndeski or GLPV in the literature. 

\subsection{DHOST theories}
Beyond Horndeski theories turned out to be  an intermediate step on the way to a much more general class of scalar-tensor theories, based on the idea of degeneracy ensuring the absence of any extra scalar degree of freedom. 
We present these scalar-tensor theories, introduced in  \cite{Langlois:2015cwa} and now known as Degenerate Higher-Order Scalar-Tensor (DHOST) theories\footnote{This name was introduced in \cite{Achour:2016rkg}. The same theories were also dubbed "Extended Scalar-Tensor" theories in \cite{Crisostomi:2016czh}.}.

DHOST theories were constructed and fully classified up to quadratic order (in second-order derivatives) in \cite{Langlois:2015cwa} (see also \cite{Crisostomi:2016czh,Achour:2016rkg,deRham:2016wji}) and later up to cubic order in \cite{BenAchour:2016fzp}. The corresponding Lagrangians can be written in the form 
\bea
\label{action}
S[g,\phi] &=& \int d^4 x \, \sqrt{- g }
\left[ F_\2(X,\phi) \,  {}^{(4)}\!R+P(X,\phi) + Q(X,\phi)  \square\phi + \sum_{I=1}^{5}A_I(X,\phi)  L^\2_ a
\right.
\cr
&&
\left. \qquad \qquad\qquad + F_\3(X, \phi) \, G_{\mu\nu} \phi^{\mu\nu}  +   \sum_{I=1}^{10} B_I(X,\phi) L^\3_a  \right] \;,
\eea
where we have introduced five elementary Lagrangians that are quadratic in $\phi_{\mu\nu}$, namely
\bea
&& L^\2_1 = \phi_{\mu \nu} \phi^{\mu \nu} \,, \qquad
L^\2_2 =(\Box \phi)^2 \,, \qquad
L_3^\2 = (\Box \phi) \phi^{\mu} \phi_{\mu \nu} \phi^{\nu} \,,  \label{Lag_quad1} \\
&& L^\2_4 =\phi^{\mu} \phi_{\mu \rho} \phi^{\rho \nu} \phi_{\nu} \,, \qquad
L^\2_5= (\phi^{\mu} \phi_{\mu \nu} \phi^{\nu})^2\,. \label{Lag_quad2}
\eea
as well as  ten cubic Lagrangians 
\bea
&& L^\3_1=  (\Box \phi)^3  \,, \quad
L^\3_2 =  (\Box \phi)\, \phi_{\mu \nu} \phi^{\mu \nu} \,, \quad
L^\3_3= \phi_{\mu \nu}\phi^{\nu \rho} \phi^{\mu}_{\rho} \,,   \\
&& L^\3_4= \left(\Box \phi\right)^2 \phi_{\mu} \phi^{\mu \nu} \phi_{\nu} \,, \quad
L^\3_5 =  \Box \phi\, \phi_{\mu}  \phi^{\mu \nu} \phi_{\nu \rho} \phi^{\rho} \,, \quad
L^\3_6 = \phi_{\mu \nu} \phi^{\mu \nu} \phi_{\rho} \phi^{\rho \sigma} \phi_{\sigma} \,,  \qquad  \\
&& L^\3_7 = \phi_{\mu} \phi^{\mu \nu} \phi_{\nu \rho} \phi^{\rho \sigma} \phi_{\sigma} \,, \quad
L^\3_8 = \phi_{\mu}  \phi^{\mu \nu} \phi_{\nu \rho} \phi^{\rho}\, \phi_{\sigma} \phi^{\sigma \lambda} \phi_{\lambda} \,,  \qquad  \\
&& L^\3_9 = \Box \phi \left(\phi_{\mu} \phi^{\mu \nu} \phi_{\nu}\right)^2  \,, \quad
L^\3_{10} = \left(\phi_{\mu} \phi^{\mu \nu} \phi_{\nu}\right)^3 \,.
\eea
The above quadratic and cubic Lagrangians represent all the possible contractions of the second-order derivatives $\phi_{\mu\nu}$ with the metric $g_{\mu\nu}$ and the scalar field gradient $\phi_\mu$.

The action (\ref{action}) contains  19 functions of $X$ and $\phi$ but these functions cannot be chosen arbitrarily to get a DHOST theory. Except the functions $P$ and $Q$, which are arbitrary, the other functions must satisfy degeneracy conditions so that the corresponding theory contains a single scalar mode. Taking into account all these degeneracy conditions\footnote{The degeneracy conditions at quadratic order will be given explicitly and the cubic ones can be found in \cite{BenAchour:2016fzp}.}, which restrict the allowed functions $F_\2$, $A_I$,    $F_\3$ and all $B_a$, one can then classify all DHOST theories of the form (\ref{action}), whose functions must satisfy specific constraints~\cite{BenAchour:2016fzp}:
\bi
\item Purely quadratic theories (with no cubic terms, i.e. such that $F_\3$ and all $B_I$ vanish) : 7 subclasses (4 subclasses with $F_\2\neq 0$, 3 with $F_\2=0$).
\item Purely cubic theories (with no quadratic term, i.e. such that $F_\2$ and all $A_I$ vanish): 9 subclasses (2 subclasses with  $F_\3\neq 0$, 7 with $F_\3=0$).
\item Mixed quadratic and cubic theories: 25 subclasses  (out of $7\times 9=63$) are degenerate. It must be stressed that the sum of two degenerate Lagragians is not necessarily degenerate.
\ei
Some details about how DHOST theories were constructed and classified will be given in section \ref{section_DHOST}, after the notion of degeneracy is explained in detail in the next section. 

\medskip 
\noindent
{\bf Particular cases}
\smallskip

DHOST theories (\ref{generalHorn}) include as a particular case the  Horndeski theories, with the identifications
\beq
F_\2=\Gfour\,, \qquad \aa=-\ab=  2 \Gfour_{,X}\,, \qquad \ac=\ad=\ae=0\,.
\eeq
for the quadratic terms, corresponding to $L_4^{\rm H}$ in (\ref{L4}), 
and
\beq
F_\3=G_5 \,,\qquad 3B_1=-B_2=\frac32B_3=G_{5,X}\,, \qquad B_I=0\ (I=4,\dots, 10)
\eeq
for the cubic terms, corresponding to $L_5^{\rm H}$ in (\ref{L5}).

DHOST theories also include the larger class of Beyond Horndeski (GLPV) theories, with the identifications
\beq
\label{Ai_GLPV}
F_\2=\Gfour\,, \qquad \aa=-\ab= 2 \Gfour_{,X}+ X \Ffour     \,, \qquad \ac=-\ad= 2 \Ffour\,, \qquad \ae=0\,.
\eeq
for the quadratic terms, corresponding to $L_4^{\rm H}+L_4^{\rm bH}$,
and
\bea
&&F_\3=G_5 \,,\quad   3B_1=-B_2=\frac32B_3=G_{5,X}+3X F_5\,, \quad 
-2 B_4=B_5=2B_6=-B_7= 6 F_5\,,
\nonumber
\\
 && B_8=B_9=B_{10}=0\,,
\eea
for the cubic terms, corresponding to $L_5^{\rm H}+L_5^{\rm bH}$.

Interestingly, Beyond Horndeski theories, and therefore Horndeski theories, belong to the same subclass of DHOST theories, dubbed class Ia in \cite{Achour:2016rkg}, although some theories in class Ia are not Horndeski or Beyond Horndeski theories. Moreover, when matter is ignored, all the Lagrangians in  class Ia can be mapped, via field redefinition of the metric (see discussion in subsection \ref{subsection:disformal}), into a Horndeski Lagrangian.

\section{Degenerate theories}
\label{section_degenerate}
Despite the presence of higher order time derivatives in their (Euler-Lagrange) equations of motion,  {\it degenerate} scalar-tensor theories contain a single degree of freedom. This can be more easily understood by considering a simple toy model in the context of classical mechanics. We will see in the next section how the notion of degeneracy operates for covariant scalar-tensor theories. The presentation of this section mainly follows the discussion given in \cite{Langlois:2015cwa}. 

\subsection{Simple toy model}
Let us  consider  the  Lagrangian
\beq\label{L_toy}
L= \frac12a\, \ddot \phi^2+ b \, \ddot \phi \, \dot q+  \frac12 c\,  \dot q^2 +\frac 12 \dot \phi^2- V(\phi,q) \,,
\eeq
where the variable $\phi$ appears with second order time derivatives, while $q$ is a standard variable, appearing in the Lagrangian with at most first order derivatives. The coefficients  $a$, $b$ and $c$ are assumed here to be constant for simplicity. 

The two equations of motion derived from (\ref{L_toy})  are given by
\begin{eqnarray}
a \ddddot \phi-  \ddot \phi  + b \dddot q   - V_\phi & = &0\; ,\label{eq for Phi}\\
c \ddot q+b \dddot \phi+  V_q   & = & 0 \label{eq for q} \,,
\end{eqnarray}
where $V_q\equiv\partial V/\partial q$ and $V_\phi\equiv\partial V/\partial \phi$.
As expected, these equations contain higher order time derivatives if $a$ and $b$ are nonzero: the first equation is fourth order if $a\neq 0$ and both equations are third order if $a=0$ and $b\neq 0$.
 In general, these higher order terms can be associated to the existence of an  extra degree of  freedom.   However,  there are special cases where the extra degree of freedom can be avoided even if the equations of motion feature higher order time derivatives. 

To compute the number of degrees of freedom (either in  the Lagrangian or Hamiltonian frameworks), it is convenient to reformulate the theory in a way that eliminates explicit higher order time derivative in the Lagrangian. For that purpose, we simply replace $\dot \phi$ by a new variable $Q$ in (\ref{L_toy}) and  add a ``constraint"  that imposes $Q=\dot\phi$. Thus, we introduce the new Lagrangian
\beq\label{reformulated toy model}
L=\frac12 a\, \dot Q^2+ b\, \dot Q \dot q + \frac12 c\, \dot q^2   + \frac12  Q^2 - V(\phi,q)  - \lambda (Q-\dot \phi)\,,
\eeq
where $Q$ and $\lambda$ are two new variables. 

The equations of motion derived from this new Lagrangian are 
\begin{eqnarray}
&& a\ddot Q + b \ddot q  =    Q - \lambda \, , \label{eq for Q} \\
&& b \ddot Q + c \ddot q  =   -  V_q \label{eq modified for q}\,\\
&&  \dot \phi \; = \; Q \;\;\; \text{and} \;\;\; \dot \lambda  =- V_\phi   \label{extra eq}\;.
\end{eqnarray}
It is easy to check that this system of equations is equivalent to  the 
original one (\ref{eq for Phi} - \ref{eq for q}). 

We now introduce the kinetic matrix, or Hessian matrix, defined by
\begin{eqnarray}
M =\left(\frac{\partial^2L}{\partial v^A \partial v^B}\right)= \left(
 \begin{array}{cc}
 a & {b} \\
{b} & c
 \end{array}
 \right) \,.
 \label{kinetic_matrix}
 \end{eqnarray} 
 If $M$ is invertible, which corresponds to the generic case, the equations (\ref{eq for Q}-\ref{eq modified for q}) enable us to express the second derivatives $\ddot Q$ and $\ddot q$ in terms of  quantities with at most first order derivatives. Together with the equations (\ref{extra eq}), the differential system thus requires  6 initial conditions, corresponding to the initial values for $Q$, $\dot Q$, $q$, $\dot q$, $\lambda$ and $\phi$. 
We thus conclude  that  the system contains $3$ degrees of freedom,  including an extra degree of freedom due to the presence of second derivatives in the Lagrangian. Note that this remains true even if  $a=0$, provided $b\neq 0$.

\subsection{Degeneracy: eliminating the extra degree of freedom} 
If the kinetic matrix $M$ is degenerate, i.e.  if its determinant vanishes, 
\beq 
ac-b^2=0\,,
\eeq
then the system does not contain any extra degree of freedom, as we show below.

An obvious way to make $M$ degenerate is to choose $a=0$ and $b=0$. In this trivial case, all the higher  order derivatives disappear in the original Lagrangian and the system describes $2$ degrees of freedom as usual.

Let us now turn to more interesting situations where the degeneracy is nontrivial. Although the associated equations of motion  involve higher order time derivatives (up to fourth order if $a\neq 0$, third order otherwise), one can  show that the degeneracy guarantees that there is no extra degree of freedom. 

In the following, we are going to assume that $c\neq 0$. 
 Combining (\ref{eq for Q} - \ref{eq modified for q}), one gets 
 \beq
 \label{eq_lambda}
 Q+\frac{b}{c} V_q=\lambda
 \eeq
 and the second equation can be rewritten as 
 \beq
 \label{eq_x}
 c \,\ddot x +V_q=0,
 \eeq
 where we have introduced 
 \beq
 x\equiv q+\frac{b}{c} \dot\phi\,.
 \eeq
 Note that $q$ can be replaced everywhere by $x- \frac{b}{c} \dot\phi$. 
 
 Taking the time derivative of (\ref{eq_lambda}) and using the second equation in (\ref{extra eq}), one gets the second-order equation of motion
 \beq
 \label{eq_phi}
 \left(1-\frac{b^2}{c^2}V_{qq}\right)\ddot\phi+\frac{b}{c} V_{qq}\, \dot x+V_\phi=0\,.
 \eeq
 We have thus  obtained a second order system for the variables $x$ and $\phi$ (the potential $V$ and its derivatives depend on $\phi$, $\dot\phi$ and $x$ via the substitution 
$q=x-(b/c)\dot \phi$),  which means that the system  only requires $4$ initial conditions, i.e. that there are only two degrees of freedom.
Note that the system (\ref{eq_x}-\ref{eq_phi})  involves the variables $\phi$ and $x$, but not $q$. However, once the system is solved  for $\phi(t)$ and the $x(t)$,  one can recover the original 
 variables via the relations $q(t)=x(t)-b \dot\phi(t)/c$.
 
 It is also instructive to count the number of degrees of freedom via a  Hamiltonian analysis of the system, which can  be found in the Appendix. In summary,  the degeneracy of the Lagrangian implies  the presence of a primary constraint. The time evolution of this primary constraint then shows the presence of a secondary constraint. 
 These two constraints, which are second-class, eliminate one degree of freedom, which corresponds to the extra degree of freedom due to the higher derivatives. The linear dependence of the Hamiltonian on one of the momenta, which is a signature of the Ostrogradski instability, is also eliminated.

To conclude, the analysis of the Lagrangian (\ref{L_toy}) shows  that the crucial ingredient to avoid the presence of an extra degree of freedom is the degeneracy of the kinetic matrix (\ref{kinetic_matrix}). 
The degeneracy of scalar-tensor theories can be investigated in a similar way, as will be illustrated in the next section. The scalar field, with second derivatives, will be analogous to our variable $\phi(t)$, while the metric field  will play a r\^ole similar to that of the ``regular'' degrees of freedom $q(t)$. Of course, the mathematical structure will be more complicated but the essential features concerning the degeneracy of the kinetic matrix will  turn out to be  quite similar.

\subsection{Generalization to multi-variable Lagrangians}

In \cite{Motohashi:2016ftl} and \cite{Klein:2016aiq}, the previous analysis was extended  to more general Lagrangians containing several variables with accelerations, namely of the form
\beq \label{lagmm}
L(\ddot \phi^\a,\dot \phi^\a , \phi^\a ; \dot q^i, q^i) \qquad (\a=1,\cdots, n; \,\, i=1,\cdots, m)\,,
\eeq 
with $\a=1, \cdots, n$ and $i=1,\cdots, m$. A priori, one generically expects $2n+m$ degrees of freedom, i.e.   $n$ extra DOFs, for this type of Lagrangians. 

As shown in \cite{Motohashi:2016ftl}, one can eliminate these $n$ extra DOFs by requiring, first,   a degeneracy of order $n$ of the kinetic matrix, or Hessian matrix:
\beq 
K=
\begin{pmatrix}
L_{\dot Q^\a \dot Q^\b} & L_{\dot Q^\a \dot q^j} \\
L_{\dot Q^\b \dot q^i} & L_{\dot q^i \dot q^j}
\end{pmatrix} \,,
\eeq
The degeneracy of order $n$ is expressed by the condition,
\beq 
L_{\a\b} - L_{\a i} L^{ij} L_{j\b} = 0 \,, 
\eeq  
which is associated with the existence of $n$ primary constraints in the Hamiltonian formulation. 

If $n=1$, this primary condition is enough to eliminate the extra DOF, as we saw in the previous subsection. 
If $n>1$, by contrast, a secondary condition is required in order to ensure that there exist $n$ secondary constraints which, together with the $n$ primary constraints, can eliminate the $n$ extra DOFs. 
This secondary condition reads 
\bea \label{secdegcon1ai} 
 &&  L_{\dot Q^\a \dot \phi^\b}
-  L_{\dot Q^\b\dot \phi^\a}
+ L_{\dot \phi^\a \dot q^i} L^{ij}  L_{\dot q^j\dot Q^\b} 
- L_{\dot Q^\a  \dot q^i} L^{ij}
L_ {\dot q^j \dot \phi^\b}  \nonumber \\
&& 
+ L_{\dot Q^\a \dot q^i} L^{ij} L_{q^j\dot Q^\b} 
- L_{\dot Q^\a  q^i} L^{ij} L_{\dot q^j\dot Q^\b}
 \nonumber \\
&& 
+ L_{\dot Q^\a  \dot q^i} L^{ij} 
\left(L_ {\dot q^j  q^k} 
    - L_ {q^j  \dot q^k} \right)
L^{kl} L_ {\dot q^l \dot Q^\b} =0\,,
\eea
If $n=1$, one can check that this condition is automatically satisfied, as expected. 
If $m=0$, it reduces to 
\beq
 L_{\dot Q^\a \dot \phi^\b}
-  L_{\dot Q^\b\dot \phi^\a}=0\,.
\eeq

Finally, it is also possible to consider generalized point particle Lagrangians with  third or higher  order time derivatives and look for theories without Ostrogradsky ghosts, as explored in \cite{Motohashi:2017eya,Motohashi:2018pxg}. Analogous  conditions  for field theories were given in 
\cite{Crisostomi:2017aim} (in particular, the property that the primary constraint is sufficiently when $n=1$ still holds for Lorentz-invariant theories).

\section{Constructing DHOST theories}
\label{section_DHOST}
The goal of this section is to show how the Lagrangians of DHOST theories can be constructed in a systematic way, illustrated by the case of quadratic DHOST theories. The construction up to third order is more involved and is presented  in detail  in \cite{BenAchour:2016fzp}. 
This rather technical section can be skipped by the reader who is more interested by the phenomenological aspects of DHOST theories, discussed in the subsequent sections. The content of this section is mainly based on \cite{Langlois:2015cwa} and 
\cite{Achour:2016rkg}. 

\subsection{Quadratic higher-order Lagrangians}
We  consider  scalar-tensor Lagrangians  with a quadratic dependence on second derivatives of the scalar field, of the form
\beq\label{generalHorn}
S[g,\phi] \equiv \int \sqrt{-g} \left[ F(X,\phi) \, R + C^{\mu\nu\rho\sigma}\,  \nabla_\mu\!\nabla_\nu\phi \, \nabla_\rho\!\nabla_\sigma\phi \right]
\eeq
where  the tensor 
 $C^{\mu\nu\rho\sigma}$ depends only on $\phi$,  $\phi_\mu$ and the metric  $g_{\mu\nu}$.  
Given the way it is contracted in the action, one can impose, without loss of generality, the following symmetries: 
\beq
C^{\mu\nu\rho\sigma} = C^{\nu\mu\rho\sigma}= C^{\mu\nu\sigma\rho}= C^{\rho\sigma\mu\nu}\,.
\eeq 
Therefore, this tensor can always be written in the form
\begin{eqnarray}\label{family}
C^{\mu\nu,\rho\sigma} & = &  \frac{1}{2} \aa\, (g^{\mu\rho} g^{\nu\sigma} + g^{\mu\sigma} g^{\nu\rho})+
\ab \,g^{\mu\nu} g^{\rho\sigma} +\frac{1}{2} \ac\, (\phi^\mu\phi^\nu g^{\rho\sigma} +\phi^\rho\phi^\sigma g^{\mu\nu} ) 
\cr
& & \quad +   \frac{1}{4} \ad (\phi^\mu \phi^\rho g^{\nu\sigma} + \phi^\nu \phi^\rho g^{\mu\sigma} + \phi^\mu \phi^\sigma g^{\nu\rho} + \phi^\nu \phi^\sigma g^{\mu\rho} ) +  \ae\, \phi^\mu \phi^\nu \phi^\rho \phi^\sigma \label{four} \,,
\end{eqnarray}
where the $A_I$  are arbitrary functions of $\phi$ and $X$. 

As a consequence, 
 the scalar part of the Lagrangian that depends on second-order derivatives of $\phi$ can be written as
\beq
\label{S_phi}
L_\phi \equiv C^{\mu\nu\rho\sigma}\,  \nabla_\mu\!\nabla_\nu\phi \, \nabla_\rho\!\nabla_\sigma\phi   = \sum_I A_I L_I^\2\,,
\eeq
expressed in terms of  the five elementary Lagrangians quadratic in second derivatives introduced in (\ref{Lag_quad1}-\ref{Lag_quad2}).

Instead of working directly with  second order derivatives  in the Lagrangian,  it is more convenient to introduce the new variable 
\beq
\label{def_Q}
\Q_\mu \equiv \nabla_\mu\phi
\eeq
 via a constraint in the Lagrangian. Our new Lagrangian is thus given by
\begin{eqnarray}\label{newform}
S[g,\phi;\Q_\mu,\lambda^\mu] = \int \sqrt{-g} \left\{\f\,   R + C^{\mu\nu\rho\sigma} \nabla_\mu \Q_\nu \, \nabla_\rho \Q_\sigma + \lambda^\mu (\nabla_\mu\phi - \Q_\mu)\right\}\,,
\end{eqnarray}
where the tensor $C^{\mu\nu\rho\sigma}$ is now expressed in terms of $\Q_\mu$ and $\phi$.
It is straightforward to verify  that (\ref{newform}) and (\ref{generalHorn}) are  equivalent.

\subsection{Kinetic matrix}
To write the degeneracy conditions, one must identify the kinetic matrix, i.e. the analog of (\ref{kinetic_matrix}). This requires a 3+1 decomposition of spacetime in order to separate time derivatives from spatial ones. 

We thus assume the existence of a slicing of spacetime with 3-dimensional spacelike hypersurfaces. We introduce their normal unit vector $n^a$, which is time-like, and satisfies the normalization condition $n_a n^a=-1$.  This induces a three-dimensional metric, corresponding to the projection tensor on the spatial hypersurfaces,  defined by
\beq
h_{ab}\equiv g_{ab}+n_a n_b\,.
\eeq
Let us also introduce a time direction vector $t^a=\partial/\partial t$ associated with a time coordinate $t$ that labels the  spacelike hypersurfaces.
One can always decompose $t^a$ as 
\beq
t^a =N n^a +N^a,
\eeq
thus defining the lapse function $N$   and the shift vector $N^a$ orthogonal to $n^a$.

It is then convenient  to decompose   $\Q_a$ introduced in (\ref{def_Q})
 into its normal and spatial projections, defined by
 \beq
\An\equiv \Q_a n^a\,,\qquad  \tA_a\equiv h_a^b\, \Q_b\,.
 \eeq
We find that
  the only terms in $\nabla_{(a}\Q_{b)}$  that contain time derivatives and are thus relevant for the kinetic part of the Lagrangian reduce to
\beq
\label{DaQb_kin}
(\nabla_{(a}\Q_{b)})_{\rm kin}=\lambda_{ab}\, \dot\An+\Lambda_{ab}^{\ \ cd} \, K_{cd}\,,
\eeq
where $K_{ab}$ is the extrinsic curvature tensor, defined by\footnote{The time derivative of any  tensor corresponds to the Lie derivative of this tensor with respect to $t^a$.}
\beq
\label{Kab}
K_{ab}\equiv \frac{1}{2N}\left(\dot h_{ab}-D_aN_b -D_bN_a\right)\,,
\eeq
and
we have introduced the tensors
\beq
 \lambda_{ab}\equiv \frac1N n_a n_b\,, \qquad \Lambda_{ab}^{\ \ cd}\equiv -\An \, h_{(a}^c h_{b)}^d+2\, n_{(a} h_{b)}^{(c} \tA^{d)}\,.
 \eeq
 Strictly speaking, only the time derivative $\dot h_{ab}$  in (\ref{Kab}) is relevant for the kinetic part of the action  but we keep $K_{ab}$ for convenience. 
 
 Substituting  (\ref{DaQb_kin}) into the action (\ref{newform}), one obtains the kinetic part of the Lagrangian, which can be written in the form
  \beq
 L_{\rm kin}=\A\, \dot\An^2+2 {\cal B}^{ab}\, \dot\An  K_{ab}
  + {\cal C}^{ab,cd}K_{ab}K_{cd}\,.
 \eeq
  This is the analog of  the Lagrangian (\ref{reformulated toy model}), with $\An$ in place of $Q$ and $K_{ab}$ (or $\dot h_{ab}$)  in the r\^ole of  $\dot q$. 
 The coefficients, analogous to  $a$, $b$ and  $c$ in (\ref{reformulated toy model}), are given by
 \beq
 \label{A}
\A\equiv C^{ef,gh}\lambda_{ef}\lambda_{gh}  \,,
 \eeq
 \beq
 {\cal B}^{ab}\equiv 2\f_X\frac{\An}{N} \, h^{ab}+C^{ef,gh}\Lambda_{ef}^{\ \ ab}\lambda_{gh}
 \label{B_decomp}
 \eeq
 and
  \begin{eqnarray}
 {\cal C}^{ab,cd} &\equiv& \frac12 \f\left(h^{ac}h^{bd}+h^{ad}h^{bc}-2h^{ab}h^{cd} \right)
 +2 \f_X\left(\tA^a\tA^b h^{cd}+\tA^c \tA^d h^{ab}\right)
 \cr
 &&
 +C^{ef,gh}\Lambda_{ef}^{\ \ ab}\Lambda_{gh}^{\ \ cd}\,.
 \end{eqnarray}
 It is worth stressing that   the scalar curvature term $\f \, \R$ gives contributions to the coefficients  ${\cal B}^{ab}$ and ${\cal C}^{ab,cd}$. For this reason,   the scalar curvature term can be considered as part of the terms at quadratic order  in $\phi_{\mu\nu}$. 
 
Interestingly, in the case of the Horndeski Lagrangian $L_4^{\rm H}$, there is a cancellation between the Ricci term  contribution and that  from the terms quadratic in $\phi_{\mu\nu}$ so that  the total coefficient ${\cal B}^{ab}$ vanishes. This is not surprising since Horndeski theories are, by construction,  restricted to give second order equations of motion. By contrast, when ${\cal B}^{ab}\neq 0$, the equations of motion become higher order, as in the case of beyond Horndeski or other DHOST theories. 

\subsection{Degeneracy conditions}

Following the simple example of the previous section, we can now derive the degeneracy conditions by considering the kinetic matrix
\beq
\left(
\begin{array}{cc}
\A &\B^{cd}\\
\B^{ab} & {\cal C}^{ab,cd}
\end{array}
\right)\,.
\eeq
 This matrix is degenerate if there exists an eigenvector with zero eigenvalue, i.e. if  one can find $v_0$ and $\V_{cd}$ such that 
\beq
\label{degeneracy}
v_0 \, \A+\B^{cd} \V_{cd}=0\,,  \,\qquad  v_0 \, \B^{ab}+ {\cal C}^{ab,cd} \, \V_{cd}=0\,,
\eeq
where  $\V_{cd}$ is a symmetric tensor of order 2. Note that the latter  can only be of the form 
\beq
\V_{cd}=v_1 \, h_{cd}+ v_2\,  \tA_c\, \tA_d\,,
\eeq
and the contraction of ${\cal C}^{ab,cd}$ with $\V_{cd}$ can be similarly decomposed along $h_{cd}$ and $\tA_c\tA_d$. In this way, the kinetic matrix can be seen as a $3\times 3$ matrix. 
Requiring the determinant of the kinetic matrix to vanish yields (using $\tA^a\tA_a=X+\An^2$)  an expression of the form 
\beq
\label{determinant}
\D_0(X)+\D_1(X)\,  \An^2+\D_2(X) \, \An^4=0\,,
\eeq
with
\beq
\label{D0}
\D_0(X)\equiv -4 (\aa+\ab) \left[X \f (2\ab+X\ad+4\f_X)-2\f^2-8X^2\f_X^2\right]\,,
\eeq
\begin{eqnarray}
\D_1(X)&\equiv& 4\left[X^2\ab (3\aa+\ab)-2\f^2-4X\f \aa\right]\ad +4 X^2\f(\aa+\ab)\ae 
\cr
&&
+8X\ab^3-4(\f+4X\f_X-6X\aa)\ab^2 -16(\f+5X \f_X)\aa \ab
\cr
&&
+4X(3\f-4X \f_X) \ab\ac 
-X^2\f \ac^2 +32 \f_X(\f+2X \f_X) \aa
\cr
&&
-16\f \f_X \ab-8\f (\f-X\f_X)\ac+48\f \f_X^2 \,,
\end{eqnarray}
\begin{eqnarray}
\D_2(X)&\equiv& 4\left[ 2\f^2+4X\f \aa-X^2\ab(3\aa+\ab)\right]\ae  +4(2\aa-X\ac-4\f_X)\ab^2
\cr
&&
+ 4\ab^3+3X^2 \ab\ac^2
-4X\f \ac^2+8 (\f+X\f_X)\ab\ac -32 \f_X \aa\ab
\cr
&&
+16\f_X^2\ab
+32\f_X^2\aa-16\f\f_X\ac\,.
\end{eqnarray}
Since the determinant must vanish for arbitrary values of $\An$, we deduce that degenerate theories are characterized by the three conditions
\beq
\D_0(X)=0, \qquad \D_1(X)=0, \qquad \D_2(X)=0\,.
\eeq

\medskip 
Note that,  if one ignores the dynamics of gravity,  the system depends only the scalar field variables and is  degenerate when $\A=0$. Using the explicit expression for $\A$,
\beq
\A=\frac{1}{N^2}\left[\aa+\ab-(\ac+\ad)\, \An^2+\ae \, \An^4\right]\,,
\eeq
 which must vanish for any value of  $\An$, one finds  the three conditions
\beq
\label{no-ghost}
\aa+\ab=0\,,  \qquad \ac+\ad=0\,, \qquad  \ae=0\,.
\eeq
One can note that both quartic Horndeski and beyond Horndeski Lagrangians satisfy these  three conditions. 

\medskip
The property that the degeneracy of the Lagrangian implies the absence of any extra scalar degree of freedom has been shown explicity in  the general case of quadratic DHOST theories by resorting to a Hamiltonian formulation that uses the variable $\An$ and does not assume any particular gauge~\cite{Langlois:2015skt}.
Earlier Hamiltonian analyses were proposed  for quadratic Beyond Horndeski theories described in the unitary gauge in \cite{Gleyzes:2014dya,Gleyzes:2014qga,Lin:2014jga} and  for a particular Beyond Horndeski model in an arbitrary gauge in \cite{Deffayet:2015qwa}. Note that there exists a special class of theories that appear degenerate in the unitary gauge but not in an arbitrary gauge (and therefore are not DHOST theories) and, in this case, the extra mode can be tamed via appropriate boundary conditions~\cite{DeFelice:2018mkq}.
Another possibility to obtain healthy scalar-tensor theories with higher derivatives  is to give up general covariance and work directly with spatially covariant theories~\cite{Gao:2014soa,Gao:2014fra}.

\subsection{Classification of quadratic DHOST theories}
\label{Section_classification}
In this subsection, we summarize the classification of quadratic DHOST theories, which can be found in detail in \cite{Langlois:2015cwa,Crisostomi:2016czh,Achour:2016rkg}. This classification has been extended to cubic order in \cite{BenAchour:2016fzp}.

For quadratic DHOST theories, the  condition $D_0(X)=0$ is the simplest of all three and allows to distinguish several  classes of theories. Indeed,  $D_0$ can vanish either if 
 $\aa+\ab=0$, which defines our first class of solutions, or  if the  term between brackets in (\ref{D0}) vanishes, which defines our second class, as well as our third class corresponding to the special case where $\f=0$. Below, we present explicitly only the first subclass of quadratic DHOST theories, which contains Horndeski and Beyond Horndeski theories, as this is the only class which turns out to be viable for phenomenology as we will explain later.

\subsubsection{Class I}
This class is characterized by $\ab=-\aa$.
\label{subsectionA}  
\bi
\item Subclass Ia (or N-I): $\f\neq X\aa$

One can  use the conditions  $D_1(X)=0$  and  $D_2(X)=0$ to express, respectively, $\ad$  and $\ae$ in terms of $\ab$ and $\ac$, provided $\f+X\ab\neq 0$. This defines the subclass Ia, characterized by
\begin{eqnarray}
\label{a4_A}
\ad&=&\frac{1}{8(\f-X\aa)^2}\left[-16 X \aa^3+4 (3\f+16 X\f_X)\aa^2
-X^2\f \ac^2\qquad 
\right.
\cr
&&\qquad 
\left.
-(16X^2 \f_X-12X\f) \ac\aa
-16 \f_X(3\f+4X\f_X)\aa
\right.
\cr
&&\qquad 
\left.
+8\f (X\f_X-\f)\ac+48\f \f_X^2\right]
\end{eqnarray}
and
\beq
\label{a5_A}
\ae=\frac{\left(4\f_X-2\aa+X\ac\right)\left(-2\aa^2-3X\aa\ac+4\f_X \aa+4\f \ac\right)}{8(\f-X\aa)^2}\,.
\eeq
Degenerate theories in class Ia thus depend on three arbitrary functions $\aa$, $\ac$ and $\f$.

\item Subclass Ib (or N-II): $\f=X\aa$

In this subclass, $ \ac=2\left(\f-2X \f_X\right)/X^2$, and 
where $\f$, $\ad$ and $\ae$ are arbitrary functions. The metric  sector is degenerate\footnote{In addition to the degeneracy of the kinetic matrix for the scalar modes, there could also be a degeneracy in the tensor part of the kinetic matrix, in which case one would lose one or more of the usual tensors modes. This possibility is due to the presence of extra primary constraints in the system. This point is discussed in \cite{Achour:2016rkg}.}.
\ei

\subsubsection{Class II}
This class is characterized by $\f\neq 0$ and $\ab\neq -\aa$.
\bi
\item Subclass IIa (or N-IIIi): $\f\neq X\aa$

Three arbitrary functions: $\f$, $\aa$ and $\ab$

\item  Subclass IIb (or N-IIIii): $\f=X\aa$

Three arbitrary functions:  $\aa$ (or $\f$),   $\ab$ and $\ac$.
 Like class Ib,  the metric sector is degenerate. 

\ei

\subsubsection{Class III}
This class is characterized by $\f=0$.

\bi

\item Subclass IIIa (or M-I): $\aa+3 \ab\neq 0$

Three arbitrary functions: $\aa$, $\ab$ and $\ac$. Note that this subclass has a non empty intersection with class Ia, which is parametrized by  i two arbitrary functions, $\aa$ and $\ac$. This intersection includes the Lagrangian $L_4^{\rm bh}$ (for which $\aa/X=\ac/2=F_4$). 
   
\item    Subclass IIIb (or M-II): $\aa+3\ab= 0$ 

Three arbitrary functions: $\ac$, $\ad$ and $\ae$, and, in general, a degenerate metric sector. 
Another special case  corresponds to the class 
\beq
\f=0\,,\qquad \aa=0\,,   \qquad {\rm (class\  IIIc)}
\eeq
which depends on four arbitrary functions. Since $\f-\aa X=0$, this class  is also degenerate in the metric sector.

\item Subclass IIIc (or M-III): $\aa= 0$

Four arbitrary functions. The metric sector is also degenerate since $\f=X\aa=0$. 

\ei

\subsection{Disformal transformations}
\label{subsection:disformal}
\def\tf{\tilde f}
\def\tg{\tilde g}
\def\tnabla{\tilde \nabla}
\def\T{{\cal T}}
\def\tf{\tilde f}
\def\ta{\tilde\alpha}
\def\r{{\gamma}}
\def\tX{\tilde X}
\def\l{\lambda}

The higher-order Lagrangians that we have considered can always be modified by a field redefinition, in particular via disformal transformations of the metric~\cite{Bekenstein:1992pj}
\beq
\label{disformal}
\tg_{\mu\nu}=C(X, \phi) g_{\mu\nu}+D(X, \phi) \, \phi_\mu\, \phi_\nu\,.
\eeq
Via this transformation,  any action $\tilde S$ written as a functional  of  $\tg_{\mu\nu}$ and $\phi$ gives  a new   action $S$  for  $g_{\mu\nu}$ and $\phi$, when one substitutes the above expression for  $\tg_{\mu\nu}$ into $\tilde S$: 
 \beq
S[\phi, g_{\mu\nu}]\equiv\tilde S\left[\phi, \tg_{\mu\nu}=C \,g_{\mu\nu}+D \, \phi_\mu\phi_\nu\right]\,.
\eeq
In this case,  the actions $S$ and $\tilde{S}$ are said to be related by the disformal transformation (\ref{disformal}).

If matter is ignored, the actions $S$ and $\tilde{S}$ describe the same physics, provided the  disformal transformation is invertible (see e.g. \cite{Domenech:2015tca} for a discussion on this point). However, if matter is included, such that the total actions read
\beq
S_{\rm tot}=S+S_m[\psi_m, g_{\mu\nu}]\,, \qquad  \tilde S_{\rm tot}=\tilde S+S_m[\psi_m, \tilde g_{\mu\nu}]
\eeq
 correspond to different physical systems, since matter is minimally coupled to the metric $g_{\mu\nu}$ in the first case and to $\tilde g_{\mu\nu}$ in the latter case\footnote{By contrast, $\tilde S+S_m[\psi_m, \tilde g_{\mu\nu}]$ and $S+S_m[\psi_m, C \,g_{\mu\nu}+D \, \phi_\mu\phi_\nu]$ describe the same physics, the matter being non minimally coupled to $g_{\mu\nu}$.}.

Interestingly, particular   subsets of DHOST theories have some special properties under restricted disformal transformations: 
\bi
\item
The family of Horndeski theories is stable under $X$-independent disformal transformations, i.e.  with $C=C(\phi)$ and $D=D(\phi)$, as shown in \cite{Bettoni:2013diz}.
\item
The family of Beyond Horndeski theories is stable under disformal transformations with $C=C(\phi)$ and $D=D(\phi,X)$, as demonstrated  in \cite{Gleyzes:2014qga}.
\item
The family of DHOST theories is stable under the most general disformal transformations~\cite{Achour:2016rkg}.
\ei

In particular, it has been checked explicitly that the structure of quadratic DHOST is preserved and that all seven subclasses are separately stable.  
Writing
\beq
\tilde S=\int d^4x\sqrt{-\tg}\left[\, {\tilde F}\, \tilde R+\sum_I{\tilde A}_I \tilde L_I^\2\right]\,,
\eeq
the relations between the new and old Lagrangians have been obtained explicitly in \cite{Achour:2016rkg}, with in particular 
\beq
\tilde L^\2_I=\sum_J\T_{IJ} L^\2_J +(\dots)\,,
\eeq
where 
$\T_{ab}$ is an upper right triangular matrix and the dots indicate terms that are at most linear in $\phi_{\mu\nu}$ and therefore not relevant for the degeneracy.

Finally, let us note that mimetic theories~\cite{Chamseddine:2013kea} and their extensions (see e.g. \cite{Sebastiani:2016ras} for a review), which are constructed via a non-invertible conformal or disformal transformation, also belong to the family of DHOST theories (even beyond the quadratic and cubic Lagrangians mentioned earlier), as discussed in \cite{Takahashi:2017pje,Langlois:2018jdg}.

\section{Effective theory of dark energy and modified gravity in cosmology}
\label{section_cosmology}
\newcommand{\alphaM}{\alpha_{\rm M}}
\newcommand{\alphaK}{\alpha_{\rm K}}
\newcommand{\alphaL}{\alpha_{\rm L}}
\newcommand{\alphaH}{\alpha_{\rm H}}
\newcommand{\alphaT}{\alpha_{\rm T}}
\newcommand{\alphaB}{\alpha_{\rm B}}
\newcommand{\bN}{\bar{N}}
\newcommand{\F}{{\cal F}}
\newcommand{\Mp}{M_{\rm P}}
\newcommand{\tzeta}{\tilde \zeta}

In this section, we study the phenomenology of DHOST theories in a cosmological context. To do so, it is very useful to resort to the unified formalism providing  an  effective description  of Dark Energy and Modified Gravity, which has been developed in a series of papers, especially \cite{Gubitosi:2012hu,Gleyzes:2013ooa,Gleyzes:2014rba,Langlois:2017mxy} (see also    \cite{Bloomfield:2012ff,Bloomfield:2013efa} and \cite{Baker:2012zs,Battye:2012eu} for related or alternative approaches).
This formalism is based on  a $3+1$  ADM decomposition of spacetime, 
\beq
\label{ADM}
ds^2=-N^2 dt^2 +h_{ij} (dx^i+N^i dt)(dx^j+N^jdt)\,,
\eeq
 in which the spatial slices coincide with uniform scalar field hypersurfaces (implicitly assuming that the scalar field gradient is time-like, which is natural in the cosmological context). In this specific gauge, often called unitary gauge, the action of DHOST theories can be written in  the form \cite{Gleyzes:2013ooa,Langlois:2017mxy}
\beq
\label{S_ADM}
S= \int d^3x \,  dt \,  N\sqrt{h}\,  L[N, K_{ij}, \R_{ij};t]\,,
\eeq
where $N$ is the lapse function which appears in  the  metric (\ref{ADM}),  $K_{ij}$ is the extrinsic curvature tensor,
 \beq
K_{ij}\equiv \frac{1}{2N}\left(\dot h_{ij}-D_i N_j -D_jN_i\right)\,,
\eeq
and $\R_{ij}$ the Ricci tensor associated with the spatial metric $h_{ij}$. 

\subsection{Homogeneous equations}
The Friedmann equations associated with  a spatially   flat Friedmann-Lema\^itre-Robertson-Walker (FLRW) geometry,
\beq
 ds^2 = - \bar{N}^2(t) dt^2 + a^2(t) \delta_{ij} dx^i dx^j\,,
 \eeq
  are then  derived from the (minisuperspace) homogeneous action
\beq
\bar S= \int   dt \,  \bar N a^3 \bar L\,,
\eeq
which is simply obtained from the general action (\ref{S_ADM}) via the substitution
\beq
\bar{L}\equiv
L[N=\bar{N}(t), K^i_j=\frac{\dot a}{\bar{N}a}\delta^i_j, \R_{ij}=0;t]\,.
\eeq

One must also take into account the action describing matter, which is assumed to be minimally coupled to the metric.  The variation of the homogeneous matter action with respect to the lapse and the scale factor define the energy density $\rho_{\rm m}$ and the pressure $p_{\rm m}$, respectively, according to the expression
\be
\label{linearmat}
\delta \bar S_{\rm m}=\int d^4x  \bN a^3\left(-\rho_{\rm m}\frac{\delta \bN}{\bN}+3 p_{\rm m}\frac{\delta a}{a}\right) \;.
\ee
Finally, the variation of the total homogeneous action $\bar{S}_{\rm total}=\bar{S}+\bar{S}_{\rm m}$ with respect to $N$ and $a$ yields the generalized Friedmann equations.
Note that, whereas there is no explicit dependence on the derivative of $N$ for Horndeski and Beyond Horndeski theories, this is no longer true for DHOST theories in general.

 \medskip
As an illustration, let us concentrate  on the subclass Ia of DHOST theories (which will  turn out to be the only viable subclass  as shown later), following the recent analysis given in \cite{Crisostomi:2018bsp}.
 Considering only terms up to quadratic order in second derivatives, one finds that the homogeneous action can be written, taking into account the degeneracy conditions, in the form 
\bea
\bar S \; &=& \; \int dt  \, N {a}^3 \left\{
-6 (F_\2 - X A_1) \left[ \frac{\dot {a}}{N a} - {\cal V} \frac{\dot\phi}{N^2} \frac{d}{dt}\left(\frac{\dot\phi}{N}\right) \right]^2 \right.
\nonumber
\\
&& \qquad \qquad \qquad  \left. -3 (Q+2F_{\2\phi}) \frac{\dot {a}\, \dot\phi}{N^2a}  - Q  \frac{1}{N} \frac{d}{dt}\left(\frac{\dot\phi}{N}\right)+ P \right\}\, ,
\label{cosmodhost}
\eea
where  ${\cal V}$ is given by
\bea
{\cal V} \; \equiv \; 
\frac{4 F_{\2 X}  +X A_3 -2A_1}{4(F_\2 - X A_1)}\, .\label{Kev}
\eea
This action must be supplemented by the matter action $S_m[\psi_m, a, N]$. By varying the total action with respect to $N$, $a$ and $\phi$, and using (\ref{linearmat}),
one obtains, respectively, the two Friedmann equations and the scalar equation of motion. 

Note that these equations contain higher order time derivatives in general. 
However, due to the degeneracy of the Lagrangian, the equations of motion can be rewritten in a system which is not higher order. The degeneracy is manifest  in the  property that the terms quadratic in $\dot a$ and $\ddot \phi$ combine into a square term in the first line of \eqref{cosmodhost}. This suggests to introduce a new scale factor that absorbs the $\ddot\phi$ dependence in the square term, i.e. such that 
\beq
\label{dot_b}
\frac{\dot b}{b}=\frac{\dot {a}}{a} - {\cal V} \frac{\dot\phi}{N} \frac{d}{dt}\left(\frac{\dot\phi}{N}\right)+\dots \, .
\eeq
 This can be achieved by defining 
\beq
a=e^{\lambda({X},\phi)} \, b \qquad \Rightarrow \qquad \frac{\dot a}{a}=\frac{\dot {b}}{b} +{\lambda_X}\dot{X}+ {\lambda_\phi}\dot\phi\,,
\label{HtoHb}
\eeq
such   that
\beq
\label{Lambda_X}
\lambda_X=-\frac12 {\cal V} = -\frac{4 F_{\2 X}  +X A_3 -2A_1}{8(F_\2 - X A_1)}\, .
\eeq
In terms of this new scale factor $b$,  the cosmological action (\ref{cosmodhost}) becomes 
\bea
\bar S=  \int dt \, N  b^3 \left[ \hat F_\2\frac{\dot{b}^2}{N^2b^2} + 
\hat Q\frac{\dot{b}\, \dot\phi}{N^2 b} + \hat P+\hat G_{(1)}\frac{1}{N} \frac{d}{dt}\left(\frac{\dot\phi}{N}\right) \right]\, ,\qquad 
\label{actionHframe}
\eea
with
\bea
&&\hat F_\2(X,\phi) \equiv -6e^{3\lambda} (F_\2-XA_1)  \, , 
 \nonumber \\
&&
\hat Q(X,\phi)  \equiv -3  e^{3\lambda} \left[ Q+2F_{\2\phi} + 4 (F_\2-XA_1) {\lambda_\phi}\right]\,, 
\\
&&
\hat G_{(1)}(X,\phi)  \equiv -  e^{3\lambda} \left[ Q+6X (Q+2F_{\2\phi}) {\lambda_X}\right]\, ,
 \nonumber \\
&&\hat P(X,\phi)  \equiv  e^{3\lambda}\left[P +3X(Q+2F_{\2\phi}) {\lambda_\phi}  + 6 X (F_\2-XA_1){\lambda_\phi^2}\right]  \,. \label{fff}
\eea
In contrast with the previous action (\ref{cosmodhost}), the variation of the total action now yields directly a second order system. Note that, since  
\beq
\frac{\delta a}{a}=-2X \lambda_X \frac{\delta N}{N}+\frac{\delta b}{b}\,,
\eeq
as follows from (\ref{HtoHb}),  one finds that 
the variation of the matter action with respect to the lapse and scale factor, given in  (\ref{linearmat}), is replaced  by 
\beq
\delta \bar S_m =b^3 \Lambda^3 \left[-\left(\rho_m +6X\lambda_X p_m\right)\delta N+ 3 p_m \frac{\delta b}{b}\right]\,.
\eeq
The properties of the homogeneous  cosmological evolution, in particular the asymptotic de Sitter regime and the matter dominated era, are discussed in 
\cite{Crisostomi:2018bsp}.

\subsection{Quadratic action}
The dynamics of linear perturbations is described by the action at quadratic order in perturbations. These perturbations are associated with the three basic ingredients of the action:
\beq
\delta N\equiv N-\bar{N}\,, \qquad \delta K^i_j=K^i_j-H\delta^i_j\,, \qquad \delta\R^i_j=\R^i_j\,,
\eeq
where $H=\dot a/(\bar{N}a)$ is the Hubble parameter, and $\R^i_j$ is already a perturbation since it vanishes in the background.
The Lagrangian at quadratic order is then obtained via a Taylor expansion, which is formally written as
\beq
L(q_A)=\bar{L}+\frac{\partial L}{\partial q_A}\delta q^A+\frac12 \frac{\partial^2 L}{\partial q_A\partial q_B}\delta q^A \delta q^B+\dots \,.
\eeq
where $q^A=\{N, K^i_j, \R^i_j\}$.

All (quadratic and cubic) DHOST theories lead to a Lagrangian quadratic in linear perturbations of the form
\begin{eqnarray}
\label{Squad}
 S_{\rm quad} &=& \int d^3x \,  dt \,  a^3  \frac{M^2}2\bigg\{ \delta K_{ij }\delta K^{ij}- \left(1+\frac23\aL\right)\delta K^2  +(1+\aT) \bigg(\delta  \R\,  \frac{\delta \sqrt{h}}{a^3} + \delta_2 \R 
 \bigg)  
 \cr
&&\qquad \qquad \qquad \qquad 
 + H^2\aK \delta N^2  +4 H \aB \delta K \delta N+ ({1+\aH}) \, \R  \, \delta N   
\cr
&&
\qquad \qquad \qquad \qquad 
+  4 \bun\,   \delta K  {\delta \dot N }   + \bdeux \, {\delta \dot N}^2 +  \frac{\btrois}{a^2}(\partial_i \delta N )^2   
\bigg\} \,, \qquad\quad 
\end{eqnarray}
where $\delta_2\R$ denotes the second order term in the perturbative expansion of $\R$, where the parameters $M$, $\aL$, $\aT$, $\aK$, $\aB$, $\aH$, $\bun$, $\bdeux$ and $\btrois$ are time-dependent functions. 

Only  the four coefficients $M$,  $\aT$, $\aK$, $\aB$ are necessary for 
Horndeski theories. Beyond Horndeski theories require in addition the parameter $\aH$. Finally, for general  DHOST theories, one must also include the  new coefficients $\aL$ and the $\beta_I$, although only one is independent since the four parameters are related by three constraints induced by the degeneracy conditions.
Let us discuss briefly these nine parameters (the first four have been introduced in \cite{Bellini:2014fua}, $\aH$ in \cite{Gleyzes:2014rba}  and the last four in \cite{Langlois:2017mxy}):
\bi
\item $\aK$: kineticity (associated with the kinetic term of the scalar field). This is the only coefficient that appears  for quintessence, k-essence.
\item $\aB$: so-called braiding parameter, corresponding to a mixing between the kinetic terms of the metric and of the scalar field\footnote{Note that the definitions of $\aB$ in \cite{Bellini:2014fua} and \cite{Gleyzes:2014rba} differ by a factor $-2$.}. 
\item $\aM$: variation rate of the effective Planck mass squared (defined as the coefficient of the kinetic terms of the tensor modes)
\beq
\aM=\frac{1}{H}\frac{d}{dt}\ln M^2\,.
\eeq

\item $\aT$: associated to the tensor propagation velocity, according to 
\beq
\label{cT}
c^2_{\rm T}=1+\aT
\eeq

\item $\aH$: this coefficient vanishes for Horndeski theories and becomes non zero for beyond Horndeski theories or DHOST theories. 

\item $\aL$: coefficient  characterizing the detuning between the $\delta K_{ij} \delta K^{ij}$ and $K^2$ terms

\item $\beta_1$, $\beta_2$ and $\beta_3$: these new coefficients  appear for  general DHOST theories, because their Lagrangian involves derivatives of the lapse function $N$, even if no extra scalar degree of freedom is involved. But they are not independent as discussed now.
\ei

\begin{figure}[h]
\begin{center}
\includegraphics[height=5.2in,width=4in, angle=-90]{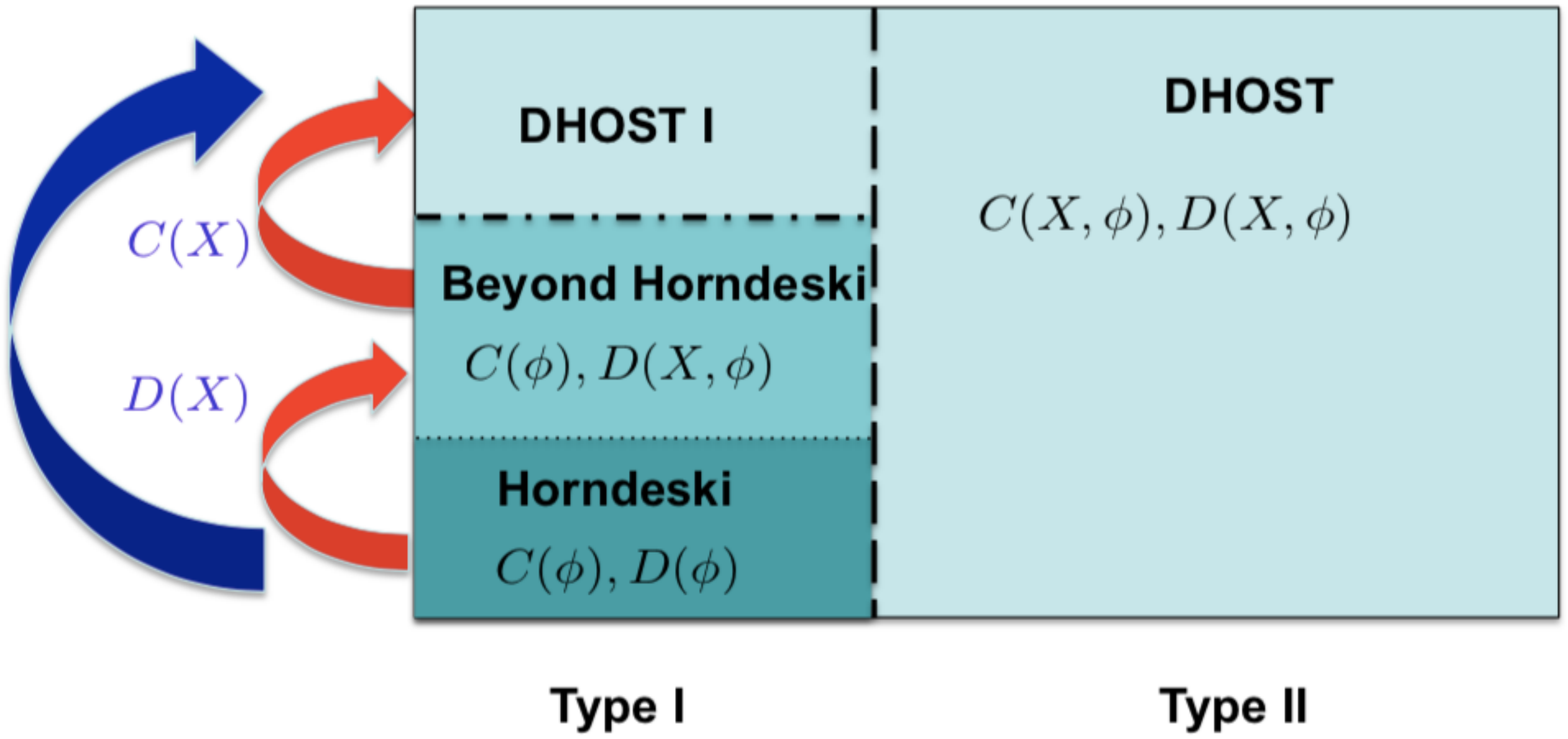}
\end{center}
\caption{\label{fig_disf_transf}
\small Type I theories (corresponding to Ia or  N-I for quadratic theories, and ${}^3\!$N-I for cubic ones, according to the terminology used in \cite{Langlois:2015cwa,Crisostomi:2016czh,Achour:2016rkg}) are related to Horndeski and Beyond Horndeski theories via disformal transformations (\ref{disformal}). By contrast, type II theories cannot be related to  Horndeski and Beyond Horndeski theories via disformal transformations.}
\end{figure}

For (quadratic and cubic) DHOST theories, the degeneracy conditions imply that the above parameters  satisfy either one of the following sets of conditions~\cite{Langlois:2017mxy}:
\beq
\label{Ia}
\CI:\quad \aL=0\,, \qquad \bdeux=-6\bun^2\,,\qquad   \btrois=-2\bun\left[2(1+\aH)+\bun (1+\aT)\right]\,,
\eeq
or 
\beq
\label{IIa}
\CII:\quad \bun=- (1+\aL)\frac{1+\aH}{1+\aT}\,, \quad \bdeux=-6(1+\aL) \frac{(1+\aH)^2 }{(1+\aT)^2}\,,\quad \beta_3=2\frac{(1+\aH)^2}{1+\aT}\,.
\eeq
The category $\CI$ contains the subclass of Horndeski theories and DHOST theories related to Horndeski via disformal transformations, i.e. the subclass Ia (see Fig. \ref{fig_disf_transf}). The category $\CII$ contains all the other subclasses.

%%%%%%%%%%%%%%%%%%%%%%%%%%%%%%
%%%%%%%%%%%%%%%%%%%%%%%%%%%%%%

\subsection{Physical degrees of freedom}
From the action (\ref{Squad}), one can extract the physical degrees of freedom, which consist of  one scalar mode and two tensor modes for DHOST theories. 

\subsubsection{Tensor modes}
At linear order, the  tensor modes correspond to the perturbations of  the spatial metric
\be
\label{h_TT}
h_{ij} = a^2(t) \left(\delta_{ij} + \gamma_{ij}\right) \;,
\ee
with $\gamma_{ij}$ traceless and divergence-free, i.e.   $\gamma_{ii}=0 = \partial_i \gamma_{ij}$. 
Substituting (\ref{h_TT}) into (\ref{Squad}), one finds that the tensor quadratic  action is given by
\beq
\label{S_quad_T}
 S_{\rm quad, tensor} = \int d^3x \,  dt \,  a^3  \bigg\{\frac{M^2}8\left[\dot\gamma_{ij}^2-\frac{1+\aT}{a^2}(\partial_k\gamma_{ij})^2\right]\bigg\} \,,
\eeq
which shows that the propagation speed of gravitational waves $c_{\rm T}$ is directly related  to the coefficient $\aT$, as announced in 
(\ref{cT}). 
The corresponding equation of motion is given, in Fourier space, by
\beq
\ddot\gamma_{ij}+\left(3\frac{\dot a}{a}+2\frac{\dot M}{M}\right)\dot\gamma_{ij}+(1+\aT)\frac{k^2}{a^2}\gamma_{ij}=0\,,
\eeq
or
\beq
\ddot\gamma_{ij}+ (3+\aM) H\dot\gamma_{ij}+(1+\aT)\frac{k^2}{a^2}\gamma_{ij}=0\,.
\eeq

\subsubsection{Scalar modes}
The scalar modes can be described 
by the 
metric
perturbations  
\be
N=1+\d N, \quad N^i=\delta^{ij}\partial_j \psi, \quad {h}_{ij}=a^2(t) e^{2\zeta}\delta_{ij}\,. \label{metric_ADM_pert}
\ee
After substitution into (\ref{Squad}), one obtains an action that depends on $\d N$, $\psi$ and $\zeta$. By solving the constraints associated with $\d N$ and $\psi$, respectively the linearized Hamiltonian and momentum constraints, one ends up with a quadratic action for the physical degree of freedom. 

\bigskip
\noindent
{\bf DHOST  with $\aL=0$ (category $\CI$)}
\smallskip

Due to the degeneracy, all the time derivatives of $\delta N$ can be combined with those of $\zeta$ through  the variable
\be
\tzeta \equiv \zeta - \bun \delta N \;.
\ee
Varying the action with respect to $\psi$ yields the scalar component of the momentum constraint. 
In terms of the new variable, it reads
\be
\label{momaz}
\delta N=\frac{\dot \tzeta}{H(1+\aB)-\dot \bun} \; .
\ee

Substituting this expression  into the action
  and performing some integration by parts,
   we finally obtain the quadratic action for the propagating degree of freedom $\tzeta$~\cite{Langlois:2017mxy}
\be
\label{actzIa}
S^\quadac= \int d^3 x \, dt \, a^3   \frac{M^2 }{2} \bigg[  {\Az}_{\tzeta}   \dot{\tzeta}^2 -  {\Bz}_{\tzeta} \frac{(\partial_i \tzeta)^2}{a^2}   \bigg] \; .
\ee
Here ${\Az}_{\tzeta}$ and  ${\Bz}_{\tzeta}$  are background-dependent 
functions  whose explicit expressions are
\begin{align}
{\Az}_{\tzeta} & = \frac{  1}{(1+\alphaB-\dot \bun /H)^2}  \bigg[\alphaK+ 6\alphaB^2 - \frac{6}{a^3 H^2 M^2} \frac{d}{dt} \left( a^3 H M^2 \alphaB \bun \right) \bigg]   \; , \label{AzI}\\
{\Bz}_{\tzeta} & =  -2 (1+\alphaT)+\frac{2}{a M^2 }\frac{d}{dt}\bigg[\frac{a M^2 \big( 1+\alphaH+\bun(1+\alphaT)\big)}{H(1+\alphaB)-\dot \bun}\bigg]   \;. \label{BzI} 
\end{align}
No instability (ghost or gradient instability) requires
\beq
{\Az}_{\tzeta}>0\,, \qquad {\Bz}_{\tzeta}>0\,.
\eeq
When $\bun=0$, one recovers the results obtained for Horndeski and Beyond Horndeski theories (see \cite{Gleyzes:2014rba}).

\bigskip
\noindent
{\bf DHOST  models in category $\CII$}
\smallskip

 For models in the category $\CII$, one finds that the coefficient of the gradient term in the quadratic action for the scalar mode reduces to~\cite{Langlois:2017mxy,Langlois:2017dyl}
  \beq
 B=-2 (1+\aT)\qquad ({\rm category}\  \CII)\,.
 \eeq
 Comparing with the tensor action (\ref{S_quad_T}), one sees that the coefficients of the gradient terms for the scalar and tensor modes have opposite signs and therefore, these modes cannot be stable simultaneously. This signals a problematic instability for all theories satisfying $\CII$, which therefore cannot be considered as viable theories.

\subsection{Cosmological constraints on DHOST models}

One finds in the literature many papers where  Horndeski models  are confronted with  cosmological data (see e.g. \cite{Barreira:2014jha,Bellini:2015wfa,Perenon:2015sla,Bellini:2015xja,Pogosian:2016pwr,Perenon:2016blf,Alonso:2016suf,Bellomo:2016xhl,Peirone:2017ywi,Kreisch:2017uet,Linder:2018jil,Reischke:2018ooh,Kennedy:2018gtx,Espejo:2018hxa} and further references in the review 
\cite{Kase:2018aps}).
Most of these papers  use either the phenomenological parameters $\mu\equiv G_{\rm matter}/G$ and $\Sigma\equiv G_{\rm light}/G$, 
or the  functions $\alpha_I(t)$ introduced in \cite{Bellini:2014fua}.  
In the latter case, one of the difficulties is to write these time-dependent functions, which in principle depends on the background evolution, in a parametrized form. Ad hoc parametrizations are mostly adopted (see discussions in e.g \cite{Linder:2016wqw} and \cite{Gleyzes:2017kpi}), with the exception of  \cite{Arai:2017hxj} where the background equations are explicitly solved. 

In the papers mentioned above, all matter species are minimally coupled to the metric. It is nevertheless possible to  relax this assumption by considering  that only ordinary matter is minimally coupled to the Horndeski metric while dark matter is non-minimally coupled. This leads to  scenarios of interacting dark energy, such as  those studied in \cite{Gleyzes:2015pma, Gleyzes:2015rua} or \cite{Amendola:2018ltt}.

Beyond Horndeski cosmology has also been considered in \cite{DeFelice:2015isa,Kase:2015zva,DAmico:2016ntq,Kase:2018iwp}, while the cosmology of  DHOST models has been explored in \cite{Crisostomi:2017pjs} and \cite{Crisostomi:2018bsp}.
Finally, let us mention that for some models, cosmological perturbations have  been studied not only at the linear level but also at the nonlinear level. In particular, the matter bispectrum has been studied for Horndeski theories \cite{Bellini:2015wfa,Cusin:2017wjg} and Beyond Horndeski theories~\cite{Hirano:2018uar}. 

\section{Astrophysical bodies in DHOST theories}
\label{section_gravity}
This section summarizes the gravitational properties of astrophysical bodies in the context of DHOST theories. 
For higher-order scalar-tensor  theories, scalar interactions can be suppressed, and therefore standard gravity recovered, via the so-called Vainshtein screening \cite{Vainshtein:1972sx}, introduced originally in the context of massive gravity. The Vainshtein mechanism has been investigated in detail in Horndeski theories~\cite{Kimura:2011dc,Narikawa:2013pjr,Koyama:2013paa}. For Beyond Horndeski theories, \cite{Kobayashi:2014ida} found the surprising result that 
the Vainshtein mechanism works only partially in the sense that the usual gravitational law is modified {\it inside matter}. This specific deviation from standard gravity was  studied for  various astrophysical  objects in order to obtain constraints on the parameters of the Beyond Horndeski models  (see e.g. \cite{Koyama:2015oma, Saito:2015fza,Sakstein:2015zoa, Sakstein:2015aac,Jain:2015edg,Sakstein:2016ggl,Babichev:2016jom,Sakstein:2016oel}), as recently summarized in \cite{Sakstein:2017xjx}. 
More recently, the Vainshtein mechanism has  been studied for  DHOST theories~\cite{Crisostomi:2017lbg,Langlois:2017dyl,Dima:2017pwp,Bartolo:2017ibw}.

\subsection{Vainshtein mechanism}

Let us consider a nonrelativistic object, characterized by a spherically-symmetric mass density $\rho(r)$. The presence of this object induces a (small) deformation of spacetime, described by the perturbed  metric
\be
ds^2=-\left[1+2\Phi(r)\right] dt^2+\left[1-2\Psi(r)\right]\delta_{ij}\, dx^i dx^j\,,
\ee
where we have introduced the two gravitational potentials $\Phi(r)$ and $\Psi(r)$. Our goal is to determine the gravitational field, i.e. $\Phi$ and $\Psi$, generated by the mass distribution $\rho(r)$. 
The scalar field $\phi$ also acquires a perturbation due to the presence of the object and can be written as 
\be
\phi=\phi_c(t)+\chi(r)\,,
\ee
where $\phi_c(t)$ denotes the cosmological value of the scalar field far from the object. 
Here, in contrast to the metric, one needs to include the cosmological dependence of the scalar field because it can be relevant even for a local observer. However,  the time dependence of $\chi$ can be neglected if one is interested in time and spatial scales much smaller than the cosmological ones, i.e. $Hr\ll1$.

The relations between the two metric perturbations, the scalar perturbation and the mass distribution can be obtained by writing down the equations of motion for the scalar field and for the metric derived from the action (\ref{action}). 
These equations are treated perturbatively, since $\chi$ and the metric perturbations are small for a nonrelativistic object. 
However, it is important to keep the nonlinear terms in higher-order derivatives of $\chi$, which become important in the Vainshtein mechanism.

Let us give the results of these calculations in the particular case of viable quadratic DHOST theories verifying
\beq
\label{cond_A12=0}
A_1=A_2=0\,,
\eeq
i.e.  such that $c_{\rm T}=1$ (as discussed in the next section), 
following the derivation presented in \cite{Langlois:2017dyl}.
In this case, keeping  only the most nonlinear terms in higher-order derivatives of $\chi$, which are relevant in regions inside or near the central object, one obtains the following three equations, corresponding,  respectively, to the scalar field equation of motion and  the time and radial components  of the metric equations of motion:
\be
\label{EOM}
\begin{split}
& (8A_3+6A_4)x^3+2 (A_3+A_4)r x\left(6x x'+rx^{\prime 2}+r xx''\right)
\\
&\  \ +\left(12\fX-(A_3-6A_4)v^2\right) x\, y+\left(4\fX+(A_3+2A_4)v^2\right)r x\, y'
\\
&\ \ -24\fX\, x \, z - 8\fX r x\,  z'=0
  \,,  \\
& \left(4\fX+(3A_3+2A_4)v^2\right)x^2+\left(4\fX+(A_3+2A_4)v^2\right)r x x'
\\
&\ \ + 2v^2(4\fX+A_4 v^2)y-4(\f+2\fX v^2)z+2\Am=0\,,
\\
& 2\fX(x^2+rxx') +(\f+2\fX v^2)y- \f z=0\,,
\end{split}
\ee
where we have used the variables $v \equiv \dot{\phi}_c$, $x\equiv \chi'/r$, $y\equiv \Phi'/r$, $z\equiv \Psi'/r$, and 
\be
\Am (r)\equiv\frac{\M (r)}{8\pi r^3}\,, \qquad \M (r)\equiv 4\pi\int_0^r\bar r^2\rho(\bar r){\rm d}\bar r\,.
\ee
The terms $A_I$, $\f$ and $\fX$ appearing in (\ref{EOM}) are all evaluated on the background. 

One can then use the last two  equation of (\ref{EOM}) to express $y$ and $z$ in terms of $x$ and $\Am$. Substituting  these expressions into the first one gives  an equation involving only the functions $x$ and $\Am$. This equation a priori involves derivatives of $x$ but, remarkably, substituting the expression (\ref{a4_A}) for $A_4$ imposed by the degeneracy conditions
 yields an equation with no derivative of $x$, which reads
\be
\begin{split}
&x\, \Big\{\left[3(4\fX+A_3 v^2)^2v^2-4\f(12\fX+7A_3v^2)\right]\Am
\\
&\quad -(4\fX+A_3 v^2)(4\f+4\fX v^2+A_3 v^4)r\Am'
\\
&\quad +4 \f A_3 (4\f+4\fX v^2-3A_3 v^4)x^2\Big\}=0\,.
\end{split}
\ee
Concentrating on the nontrivial solution, the above expression yields $x^2$ in terms of the matter function $\Am$ and its derivative $\Am'$. Substituting back into the earlier expressions for $y$ and $z$, it is then easy to write $y$ and $z$ in terms of $\Am$ and its derivatives. This finally gives the modified gravitational laws~\cite{Langlois:2017dyl} (see also \cite{Crisostomi:2017lbg})
\be
\label{mod_gravity}
\begin{split}
\frac{d\Phi}{dr}&= \frac{\GN \M(r)}{r^2}+\Xi_1 \GN \M''(r)\,,
\\
\frac{d\Psi}{dr}&= \frac{\GN \M(r)}{r^2}+\Xi_2\frac{\GN \M'(r)}{r}+\Xi_3\,  \GN \M''(r)\,,
\end{split}
\ee
with the effective Newton's constant
\be
(8\pi\GN)^{-1}=2\f+2\fX v^2- \frac32 A_3 v^4\equiv  2\f (1+\Xi_0)
\ee
 and the dimensionless coefficients
\be
\label{Xi}
\Xi_1=-\frac{(4\fX+A_3 v^2)^2}{16\f A_3}\,,\qquad
\Xi_2=\frac{2\fX v^2}{\f}\,, \qquad  \Xi_3=\frac{16\fX^2-A_3^2 v^4}{16A_3 \f}\,.
\ee

Outside the matter source, $\M$ is constant and one recovers the usual  gravitational behaviour. But, inside the matter distribution,  the above equations differ from those of standard gravity. In the Beyond Horndeski case, we have $4\fX+X A_3=0$, and therefore, 
$\Xi_3=0$ and  $\Xi_2=-\Xi_0=-2\, \Xi_1$ in this particular case. 

Interestingly, the three coefficients in (\ref{Xi}) are not independent but satisfy the consistency relation
\be
\label{consistency}
\Xi_3^2-\Xi_1^2=\frac12 \Xi_1\, \Xi_2\,.
\ee
It is also worth noting that
the coefficients $\Xi_i$ can be related to two of the  cosmological effective parameters, namely  $\aH$ and $\beta_1$. One finds the very simple relations
\be
\Xi_1=-\frac{(\aH+\beta_1)^2}{2(\aH+2\beta_1)}\,,\qquad   \Xi_2=\aH\,, \qquad  \Xi_3=-\frac{\beta_1(\aH+\beta_1)}{2(\aH+2\beta_1)}\,.
\ee
The above expressions are valid when $\aT=0$, corresponding to $A_1=A_2=0$. In the more general situation where $\aT$ is left unconstrained, viable quadratic DHOST theories still lead to 
gravitational laws of the form (\ref{mod_gravity}) with more general  expressions for the coefficients $\Xi_I$, which  can be found in 
\cite{Dima:2017pwp}.

\subsection{Newtonian stars}
For Newtonian stars, only the first equation of (\ref{mod_gravity}) is relevant, i.e. 
\beq
\frac{d\Phi}{dr}= \frac{\GN \M(r)}{r^2}+\Xi_1 \, \GN \, \M''(r)\,.
\eeq
In the literature,  the coefficient $\Upsilon\equiv 4 \, \Xi_1$
is often used. 

A theoretical  upper bound  is  \cite{Saito:2015fza}
\beq
\label{lower_bound_Upsilon}
 \Xi_1>-\frac16\,,
\eeq
just to ensure that the mass density $\rho$ decreases when going outwards from the center. Indeed, near the center, we have
\beq
\M\approx \frac{4\pi}{3} \rho_c r^3\,, \qquad \M''\approx 6\frac{\M}{r^2}\,,
\eeq
where $\rho_c$ is the central mass density, 
and therefore 
\beq
\frac{d\Phi}{dr}\approx (1+6\, \Xi_1) \frac{\GN \M}{r^2}\,,
\eeq
which gives the above constraint in order to have $\Phi'>0$, i.e. an attractive gravitational force. 

Various astrophysical constraints on $\Xi_1$ have been given in the literature. An upper bound was obtained by considering the lowest mass hydrogen burning stars~\cite{Sakstein:2015zoa}. Indeed,  for $\Xi_1 >0$, the onset of hydrogen  burning in low mass stars occurs at higher masses than in standard gravity. The lowest mass observed ($\simeq 0.1 M_\odot$) thus gives the upper bound
\beq
\Xi_1 <0.4 \,.
\eeq
A more recent  upper bound,
\beq
\Xi_1 \leq 3.5\times 10^{-2}\, \quad  (2\sigma)\,,
\eeq
has been obtained  \cite{Saltas:2018mxc} from the observation of white dwarfs (radius, mass and temperature): one of the difficulties in using white dwarf is to take into account the finite temperature effects and the existence of an envelope. 
This result improves  a previous bound  also obtained from observations of white dwarfs in \cite{Jain:2015edg}.
The same reference found a lower bound $\Xi_1 \geq -4.5\times 10^{-2} \ (1 \sigma)$,  based on the Chandrasekhar mass limit of white dwarfs\footnote{Taking $\Xi_1<0$ decreases the Chandrasekhar mass of white dwarfs. Requiring that it  must remain higher than the mass of the most massive white dwarf ($1.37 M_\odot$) gives a lower bound on $\Xi_1$.}.
However, taking into account relativistic effects, one finds simply $\Xi_1 \geq -0.11$ \cite{Babichev:2016jom}.
Finally, analysis of weak lensing and X-ray profiles of $58$ galaxy clusters have given the constraint 
$\Upsilon=4\Xi_1=-0.11^{+0.93}_{-0.67}$~\cite{Sakstein:2016ggl}.

\subsection{Neutron stars}
Neutron stars in the context of DHOST theories have been explored in several papers, not only  within  Horndeski theories (see e.g. \cite{Maselli:2015yva,Cisterna:2016vdx,Maselli:2016gxk} and the review \cite{Babichev:2016rlq}) but also for Beyond Horndeski and DHOST theories, as we discuss below.

Neutron stars  in the context of Beyond Horndeski gravity have been studied numerically in \cite{Babichev:2016jom} and \cite{Sakstein:2016oel}
 with an action of the form
\beq
S= \int d^4 \sqrt{-g}\left[ M_P^2 \left(\frac{R}{2}-k_0 \Lambda\right)- k_2 X-\frac{\zeta}{2} X^2+f_4 L^{\rm bH}_4\right]\,,
\label{action_NS}
\eeq
where $k_0$, $k_2$, $\zeta$ and $f_4$ are assumed to be constant. 
The action is expressed in the physical (Jordan) frame where matter is minimally coupled to the metric. 
Reference \cite{Babichev:2016jom} studied the case $\zeta=0$, with a polytropic equation of state 
\beq
\label{EoS_polytrope}
\rho=\left(\frac{P}{K}\right)^{1/2}+P\,, \qquad (K=123 M_{\odot}^2)\,, 
\eeq
as well as  two realistic equations of state, called SLy4 and BSK20.  The case $\zeta\neq 0$  was considered in \cite{Sakstein:2016oel}  with a broad range of realistic equations of state. 

Since local gravity depends on the cosmological velocity of the scalar field, it is instructive to consider a star embedded in a cosmological spacetime.
The simplest cosmological solution is de Sitter, which can be written, in Schwarzschild-like coordinates, as 
\beq
\label{dS}
ds^2 =-(1-H^2 r^2) dt^2+\frac{dr^2}{1-H^2 r^2}+r^2\left(d\theta^2+\sin^2\!\theta\,  d\phi^2\right)\,,
\eeq
while the scalar field profile $\phi=v t_c$ (where $t_c$ is the usual cosmological time) becomes
\beq
\phi(r,t)=v t+\frac{v}{2H} \ln (1-H^2 r^2)\,.
\eeq
One can then insert a spherical symmetric object within this cosmological solution by trying to solve the Einstein equations for a static and spherically symmetric metric of the form  
\beq
ds^2=-e^{\nu(r)}dt^2+e^{\lambda(r)}dr^2+r^2\left(d\theta^2+\sin^2\!\theta \, d\phi^2\right)\,,
\eeq
going asymptotically to (\ref{dS}). 

In practice, the energy-momentum tensor is the sum of the scalar field  contribution and that of a perfect fluid  assumed to model the neutron star's matter. Remarkably, for theories of the form (\ref{action_NS}), one finds  neutron star solutions where the metric outside the star is  an exact Schwarzschild-de Sitter. In fact, they belong to a subset  of theories among Beyond Horndeski theories leading to Schwarzschild-de Sitter, characterized by four specific conditions verified by the six  functions  in the Lagrangian, as discussed in \cite{Babichev:2016kdt}. 

 Inside the star, the equations of motion can be solved  numerically, assuming some equation of state, in order to determine the matter density profile and the internal geometry. 
When $\Xi_1 <0$, one finds that stars with fixed mass have a larger radius than their GR counterparts. Moreover, for the same equation of state, the maximum mass can increase significantly with respect to GR~\cite{Babichev:2016jom}. This illustrates how modified gravity could provide a solution to the hyperon puzzle. 

Slowly-rotating neutron stars were also studied in \cite{Sakstein:2016oel}. Interestingly,  one can derive a relation between the dimensionless moment of inertia $Ic^2/G^2M^3$ and the compactness $GM/Rc^2$, which is robust in the sense that it weakly depends on the equation of state and which can discriminate between modified gravity and GR.

The action (\ref{action_NS}) gives models where the speed of gravitational waves differs from that of light. Such models are in contradiction with the recent observation of a neutron star merger via gravitational waves and gamma rays, as we will discuss in the next section. The most recent studies have taken into account this constraint and investigated models for which $c_g=c$. For example, a Beyond Horndeski model characterized by 
$F_4=-2 G_{4X}/X$, so that $A_1=A_2=0$ according to (\ref{Ai_GLPV}), was studied in \cite{Chagoya:2018lmv}.

Finally, a  DHOST model satisfying $c_g=c$ was considered in  \cite{Kobayashi:2018xvr}, with an action of the form 
\beq
S= \int d^4 \sqrt{-g}\left[ \f R+A_3 L_3 +A_4 L_4+A_5 L_5\right]\,,
\eeq
where 
\beq
\f=\frac{M_p^2}{2}+\alpha X^2\,, \qquad A_3=-8\alpha -\beta\,, 
\eeq
and the functions $A_4$ and $A_5$ are fixed in terms of $\f$ and $A_3$ by the degeneracy conditions.
Note that the limit  $\beta=0$ corresponds to a Beyond Horndeski model. 
Using the polytropic equation of state (\ref{EoS_polytrope}), neutron stars that differ significantly from GR ones were obtained, even for small values of 
\beq
\bar \alpha= \alpha \frac{v^4}{M_p^2}\,,\qquad \bar \beta= \beta \frac{v^4}{M_p^2}\,.
\eeq
Moreover, no qualitative difference was noticed between the cases $\beta=0$ and $\beta\neq 0$, i.e. between Beyond Horndeski  and other DHOST models. 

Black holes have been studied in DHOST theories, in particular in Horndeski and Beyond  Horndeski theories (see e.g. 
\cite{Babichev:2017guv,Babichev:2017lmw,Chagoya:2018lmv} and the review \cite{Babichev:2016rlq}).

\section{DHOST theories  after GW170817}
\label{section_GW}
The propagation of gravitational waves in a cosmological background can differ from that  of light, as illustrated by   (\ref{cT}). 
 The simultaneous detection of a gravitational wave signal and its optical counterpart has thus provided very interesting constraints on dark energy models (see e.g. \cite{Ezquiaga:2018btd} for a review and references therein). This was dramatically illustrated by the recent observation of a neutron star merger, associated to the gravitational wave event  GW170817 seen by LIGO/Virgo~\cite{TheLIGOScientific:2017qsa} and the  gamma-ray burst GRB170817A~\cite{Monitor:2017mdv}, showing in particular that the speed of gravitational waves and that of light coincide to a very high precision, namely
\beq
-3\times 10^{-15}\le \frac{c_g}{c}-1\le 7\times 10^{-16}\,.
\eeq
For simplicity here, we will ignore the possibility of some (extreme) fine-tuning and assume that this observation is a strong indication    that $c_g$ and $c$  {\it strictly} coincide. The implications of $c_g=c$ for  higher-order scalar-tensor theories, summarized in several papers (see e.g. \cite{Creminelli:2017sry,Ezquiaga:2017ekz,Sakstein:2017xjx,Baker:2017hug,Langlois:2017dyl,Crisostomi:2017lbg}), are given below, considering first the general case of DHOST theories, and then specializing to Beyond Horndeski and Horndeski theories. 

Note that these implications of GW170817 for DHOST theories considered as dark energy candidates assume  that the regime of validity of these effective theories remains valid up to an energy scale corresponding to the frequency of gravitational waves observed by LIGO/Virgo, $f\sim 100$ Hz. 
However, this is very close to the typical strong coupling scale associated with these dark energy models, $\Lambda \sim (M_PH_0^2)^{1/3}\sim 260$ Hz. As pointed in  \cite{deRham:2018red}, one could envisage that the effective description of modified gravity near this scale is such that $c_g$ is very close to $1$, even if $c_g$ can significantly deviate from $1$ on much lower energy scales. 
The restrictions on DHOST theories presented below should thus be taken with a grain of salt.

\subsection{Constraints on DHOST theories}
Since the speed of gravitational waves (with respect to a cosmological background) has been  computed for all DHOST theories, it is a straightforward exercise to identify the DHOST theories that ``survive'' after GW170817. In fact, it is more useful to consider directly the full  ADM decomposition  of the DHOST Lagrangian (in a gauge where the scalar field is spatially uniform), because it provides nonlinear information and thus makes the  background-dependent effects manifest. Concentrating on the  terms
in the Lagrangian that contribute to the kinetic and spatial gradient terms  of the tensor modes, one finds~\cite{Langlois:2017mxy}
\beq
L_{\rm ADM}= (\f - X A_1)\, K_{ij}K^{ij}+\f \,   {}^{(3)}\!R+\dots
\eeq
Note that the cubic terms  also contribute to the gravitational speed  but in an explicitly background-dependent way. The requirement $c_g=c$ for any background thus imposes the very simple condition 
\beq
\label{constraint_GW}
A_1=0
\eeq
for the quadratic terms, while all cubic terms must vanish. 

Going back to  the quadratic DHOST theories identified in \cite{Langlois:2015cwa},  one sees that the condition $c_g=c$ dramatically restricts the set of viable theories, which must now depend on only two arbitrary functions of $\phi$ and $X$, namely $\f$ and $A_3$. The other functions $A_I$ are given, according to (\ref{constraint_GW}) and (\ref{a4_A}-\ref{a5_A}), by
\beq
\label{A_cg=1}
\begin{split}
& A_1 = A_2=0\,,  \\
& A_4=\frac{1}{8\f}\left[ 48 \fX^2 -8(\f-X\fX) A_3-X^2 A_3^2\right] \,, \\
& A_5 =\frac{1}{2 \f}\left(4\fX+X A_3\right) A_3\,,
\end{split}
\eeq
which corresponds to the Lagrangian
\beq
\label{DHOST_cg=1}
\begin{split}
& L^{_{\rm DHOST}}_{c_g=1}=  \p + \q\,  \Box\phi +  \f  \, {}^{(4)}\! R +  \h\phi^\mu \phi^\nu \phi_{\mu \nu} \Box \phi  \\
&+\frac{1}{8\f} \bigg(48 \fX^2 -8(\f-X\fX) \h-X^2 \h^2 \bigg) \phi^\mu \phi_{\mu \nu} \phi_\lambda \phi^{\lambda \nu} \\
&+\frac{1}{2 \f}\left(4\fX+X \h\right) \h(\phi_\mu \phi^{\mu \nu } \phi_\nu)^2 \;. 
\end{split}
\eeq

\subsection{Constraints on Horndeski and Beyond Horndeski theories}
One can then specialize the above results to Beyond Horndeski theories, characterized by 
\beq
\f=\Gfour\,, \quad   A_1=-A_2=2 \Gfour_X+ X \Ffour\,, \quad   A_3=-A_4=2 \Ffour\,,\quad
A_5=0\,.
\eeq
Combining this with the condition $A_1=0$ combined  implies 
\beq
\Ffour=-\frac{2}{X} \Gfour_X=-\frac{2}{X}\fX\,.
\eeq 
The corresponding Lagrangian is thus obtained by substituting $A_3=-4\fX/X$ in (\ref{DHOST_cg=1}), which gives
 \beq
\label{bH_cg=1}
L^{_{\rm bH}}_{c_g=1}=  \p + \q\,  \Box\phi +  \f  \, {}^{(4)}\! R  -\frac{4}{X}\fX \left(\phi^\mu\phi_{\mu \nu}  \phi^\nu \Box \phi  
-\phi^\mu \phi_{\mu \nu} \phi_\lambda \phi^{\lambda \nu} \right)
 \;. 
\eeq

Finally, one can further specialize to the case of Horndeski theories, characterized by $F_4=0$, which implies $F_X=0$. The corresponding Lagrangian thus reduces to 
\beq
\label{bH_cg=1}
L^{_{\rm bH}}_{c_g=1}=  \p(X,\phi) + \q(X,\phi)\,  \Box\phi +  \f(\phi)  \, {}^{(4)}\! R\,.
\eeq

\section{Conclusions}

In this review, we have presented the family of  higher-order scalar-tensor theories, dubbed DHOST theories, that do not contain any problematic extra degree of freedom because of the degeneracy of their Lagrangian.  It should be pointed out that  similar analyses have been performed for vector-tensor theories (see in particular \cite{Heisenberg:2016eld,Kimura:2016rzw} and other references in the review \cite{Heisenberg:2018vsk}) and for exploring theories beyond Lovelock \cite{Crisostomi:2017ugk}.

As we have seen, from a very large family of theories, explicitly constructed up to  cubic order in second derivatives $\phi_{\mu\nu}$, emerges a subset of viable theories, safe from gradient instabilities (in both  scalar and tensor sectors), which can be disformally related to Horndeski theories (but matter cannot always be minimally coupled to the metric in the  Horndeski formulation). These theories  are in principle potential models of dark energy and can be confronted to cosmological and astrophysical observations. Remarkably, the recent observation of a binary neutron star merger, showing that the speed of gravitational waves coincides with that of light,  has put severe constraints on DHOST theories\footnote{It has been recently argued in \cite{Creminelli:2018xsv} that  more theories are in fact  ruled out as dark energy candidates  by the direct detection of gravitational waves, because they predict a rapid decay of gravitational waves into scalar field fluctuations. This
would leave as acceptable theories only the DHOST theories (\ref{DHOST_cg=1}) with the additional constraint $A_3=0$, i.e. with $L=\p + \q\,  \Box\phi +  \f  \, {}^{(4)}\! R+6 (\fX^2/\f)  \phi^\mu \phi_{\mu \nu} \phi_\lambda \phi^{\lambda \nu}$. Note that this effect assumes again that the effective description on cosmological scales remains valid on the much higher energy scale of LIGO/Virgo gravitational waves.}, assuming that they are still applicable at  energy scales probed by LIGO/Virgo.

Although we  have focused our attention  on scenarios of dark energy, it should  be mentioned that higher-order scalar-tensor  theories have also been considered in the context of early universe cosmology. In particular, Horndeski and Beyond Horndeski theories have been used to construct scenarios such as cosmological bounces or Genesis, which require a violation  of the Null Energy Condition (see e.g. \cite{Creminelli:2010ba,Ijjas:2016tpn,Libanov:2016kfc,Creminelli:2016zwa,Kolevatov:2017voe,Ijjas:2017pei,Mironov:2018oec}). 

Another issue concerning DHOST theories is the robustness of their Lagrangian with respect to quantum corrections. Galileon theories are protected by the so-called galileon symmetry $\phi\rightarrow \phi+ b_\mu x^\mu$. When gravity is taken into account, this galilean symmetry is no longer exact but it is nevertheless possible to identify theories with weakly broken galileon  invariance, which are still stable with respect to quantum corrections~\cite{Pirtskhalava:2015nla}. This analysis has been recently extended to quadratic DHOST theories~\cite{Santoni:2018rrx}, showing that theories with these properties form a subset of Beyond Horndeski theories\footnote{Two conditions on quadratic DHOST theories have been obtained in \cite{Santoni:2018rrx} by computing the loop corrections. Interestingly, the first condition turns out to be equivalent to  $\A=0$, meaning that the theory is non degenerate when gravity is ignored, while the second condition is the same as that obtained from the degeneracy conditions in the case  $\A=0$,  as mentioned in \cite{Langlois:2015cwa}.}.

A more mathematical  issue is whether the initial value problem, at least locally, is well posed. A necessary condition for local well-posedness is the strong hyperbolicity of the equations of motion in some gauge, such as Einstein's equations in the harmonic gauge.  It was shown in \cite{Papallo:2017qvl,Papallo:2017ddx} that one can  find a generalized harmonic gauge in which the equations of motion, for a generic weak-field background, are strongly hyperbolic only for a very limited subclass of Horndeski theories (satisfying $G_3=G_{4X}=G_5=0$). These results hold for the generalized harmonic gauge and there could exist, in principle, a different gauge where a larger subset of theories 
could be shown to be strongly hyperbolic.

As this review has tried to illustrate, scalar-tensor theories with higher order derivatives have revealed fascinating uncharted territories, generating  a series of new issues at both fundamental and phenomenological levels. 

\medskip

\subsection*{Acknowledgments}
The author is grateful to  all his collaborators for their crucial contributions to the topic discussed in this review, especially  Karim Noui, Filippo Vernizzi and Marco Crisostomi.  I would also like to  thank the Princeton Center for Theoretical Science for their invitation to the Workshop "Gravity in the Early Universe" (January 2018), as well as  the Yukawa Institute for Theoretical Physics at Kyoto University for their invitation to the symposium YKIS2018a ``General Relativity -- The Next Generation", where  parts of the contents of this review were presented orally.  Finally, let me thank M. Sami for inviting me to write this article that summarises the research on DHOST theories.

\medskip

\appendix
\section{Degeneracy and Hamiltonian formulation}
The Hamiltonian point of view provides  the most systematic approach to count the numbers of physical degrees of freedom and to study the stability  of a system. As a complement to the  Lagrangian formulation described in section \ref{section_degenerate}, we present here the Hamiltonian analysis, following \cite{Langlois:2015cwa}. The extension of this analysis to the case of quadratic DHOST theories can be found in 
\cite{Langlois:2015skt}.

In the Hamiltonian framework, the configuration variables and  their respective conjugate momenta satisfy the Poisson brackets:
\begin{eqnarray}
\{P,Q\} = 1 \;\;, \;\; \{p,q\}=1 \;\;,\;\; \{\pi_\phi,\phi\}=1 \,,
\end{eqnarray}
while all other Poisson brackets vanish. For simplicity,  $\lambda$ is not considered as another variable, but simply as the momentum $\pi_\phi$ since, according to (\ref{reformulated toy model}), $\pi_\phi\equiv \partial L/\partial \dot\phi=\lambda$.  One also finds that the momenta $P$ and $p_i$ are related to $\dot Q$ and $\dot q$ by 
\begin{eqnarray}
\label{Pi}
 \left( \begin{array}{c} P \\ p \end{array} \right) =\left( \begin{array}{c} \frac{\partial L}{\partial\dot Q} \\ \frac{\partial L}{\partial \dot q} \end{array} \right)= \left( \begin{array}{c} a\dot Q+b\dot q \\ b\dot Q+c\dot q \end{array} \right) = M \left( \begin{array}{c} \dot{Q} \\ \dot{q} \end{array} \right) \,,
\end{eqnarray}
where $M$ is the kinetic matrix (\ref{kinetic_matrix}).

In the nondegenerate case, one can invert  the system (\ref{Pi}) in order to express the velocities $\dot Q$ and $\dot q$ in terms of the momenta. The Hamiltonian is thus given by
\begin{eqnarray}
H & = & P\dot Q + p\dot{q} + \pi_\phi \dot \phi - L \\
& = & \frac{1}{2} \left(\, P , \ p \, \right) M^{-1} \left( \begin{array}{c} P \\ p \end{array} \right) + V(\phi,q) -\frac{1}{2}Q^2 + \pi_\phi Q \,.
\end{eqnarray}
The Hamiltonian $H$ is a function of the $6$ canonical variables $(Q,q,\phi)$ and $(P,p,\pi_\phi)$, corresponding to $3$ degrees of freedom, in agreement with the results of  the Lagrangian analysis. Moreover,  one observes that the Lagrangian is linear in $\pi_\phi$, which makes the Hamiltonian unbounded from below.  This is the characteristic signature of Ostrogradski's instability~\cite{Woodard:2015zca}.

In the degenerate case, one cannot express the velocities $\dot Q$ and $\dot q$  in terms of the momenta. Instead, one finds
 a primary constraint  relating the canonical momenta, given by 
\begin{eqnarray}
\Omega =  \frac{b}{c}p - P \, \approx \, 0.
\end{eqnarray}
As usual, we use the notation $\approx$  to denote  weak equality in phase space. Taking into account this constraint, the total Hamiltonian  is given by
\begin{eqnarray}\label{def Hamilton}
H_{\rm tot} = P\dot Q+p \dot q + \pi_\phi \dot \phi- L + \mu \,\Omega
\end{eqnarray}
where $\mu$ is a Lagrange multiplier enforcing the primary constraint. 

After some straightforward manipulations to eliminate the velocities, one finds that the expression of $H_{\rm tot}$  in terms of the canonical variables reduces to
\begin{eqnarray}
H_{\rm tot} = \frac{1}{2c} p^2 - \frac{1}{2}  Q^2 + V(\phi,q) + \pi_\phi Q + (\mu+\dot Q)  \Omega \,.
\end{eqnarray}
One can then redefine the Lagrange multiplier $\mu$ so that  $\mu + \dot Q$ becomes $\mu$. 

The invariance under time evolution of the constraint $\Omega$ leads to the secondary constraint
\begin{eqnarray}
\Psi = \dot \Omega=\{\Omega , H_{\rm tot}\} = Q + \frac{b}{c} V_q - \pi_\phi \; \approx \; 0 \,.
\end{eqnarray}
To see whether the time evolution of $\Psi$ leads to a tertiary constraint, it is sufficient to compute the Poisson bracket between the 
primary and secondary constraints
\begin{eqnarray}
\{ \Omega , \Psi\} = 1- \frac{b^2}{c^2} V_{qq} \equiv  \Delta \;,
\end{eqnarray}
Leaving  aside the special case $\Delta=0$ (which would generate new constraints and further reduce the physical number of degrees of freedom), 
the analysis stops in the generic case where $\Delta\neq 0$,  because imposing $\dot\Psi=0$ simply fixes the Lagrange multiplier $\mu$ without generating any new constraint. 

In summary, we have  obtained a Hamiltonian system in a $6$-dimensional phase space, restricted by two second-class constraints $\Omega$ and $\Psi$. This implies that the number of physical degrees of freedom is only $2$. This confirms that there is no extra degree of freedom in the degenerate case. 

Moreover, one can construct  the physical phase space spanned by the variables $(q,p;Q,\phi)$, with  a Poisson algebra defined from the Dirac bracket
\begin{eqnarray}
\{F,G\}_D = \{F,G\} - \frac{1}{\Delta} \left( \{F,\Psi\}\{\Omega,G\}  -\{F,\Omega\}\{\Psi,G\}  \right) \,.
\end{eqnarray}
The associated  Hamiltonian $H_{\rm phys}$ is obtained from the total Hamiltonian after elimination of 
  $P$ and $\pi_\lambda$ via the second class constraints: 
\begin{eqnarray}
H_{\rm phys} = \frac{1}{2c} p^2+ \frac{1}{2} Q^2 + \frac{b}{c} V_qQ  + V(\phi,q) \,.
\end{eqnarray}
The linear dependence on a canonical momentum has disappeared in the above Hamiltonian. This confirms that the  Ostrogradski ghost has disappeared as a consequence of the degeneracy of the kinetic matrix $M$.

\bibliographystyle{utphys}
\bibliography{/Users/langlois/Documents/arts/18IJMPD/Dark_Energy_biblio_1809}
\end{document}